\documentclass[conference]{IEEEtran}
\IEEEoverridecommandlockouts
\usepackage{cite}
\usepackage{color}
\usepackage{amsmath,amssymb,amsfonts}
\usepackage{algorithmic}
\usepackage{graphicx}
\usepackage{textcomp}

\usepackage{xcolor}
\usepackage{colortbl}

\def\BibTeX{{\rm B\kern-.05em{\sc i\kern-.025em b}\kern-.08em
    T\kern-.1667em\lower.7ex\hbox{E}\kern-.125emX}}

\usepackage[ruled]{algorithm2e}

\usepackage{url}

\def\BibTeX{{\rm B\kern-.05em{\sc i\kern-.025em b}\kern-.08em T\kern-.1667em\lower.7ex\hbox{E}\kern-.125emX}}

\newcommand {\mymarginpar}[1]{\marginpar{#1}}
\renewcommand {\marginpar}[1]{}

\def\_{\rule{.3em}{.15ex}}      

\newcommand{\ls}[1]
   {\dimen0=\fontdimen6\the\font
    \lineskip=#1\dimen0
    \advance\lineskip.5\fontdimen5\the\font
    \advance\lineskip-\dimen0
    \lineskiplimit=.9\lineskip
    \baselineskip=\lineskip
    \advance\baselineskip\dimen0
    \normallineskip\lineskip
    \normallineskiplimit\lineskiplimit
    \normalbaselineskip\baselineskip
    \ignorespaces
   }

\newcommand {\bearn}{\begin{eqnarray*}}
\newcommand {\eearn}{\end{eqnarray*}}
\newcommand {\barr}{\begin{array}}
\newcommand {\earr}{\end{array}}

\newcommand {\N}{{\cal N}}

\newtheorem{definition}{Definition}
\newtheorem{property}[definition]{Property}
\newtheorem{proposition}[definition]{Proposition}
\newtheorem{lemma}[definition]{Lemma}
\newtheorem{theorem}[definition]{Theorem}
\newtheorem{corollary}[definition]{Corollary}
\newtheorem{example}{Example}
\newtheorem{remark}[definition]{Remark}

\newcommand{\comb}[2]
{\left ( \begin{array}{c} #1 \\#2 \end{array} \right ) }

\newcommand {\benum} {\begin{enumerate}}
\newcommand {\eenum} {\end{enumerate}}

\newcommand {\bdesc} {\begin{description}}
\newcommand {\edesc} {\end{description}}

\newcommand {\bfig}[2] {\begin{figure}
  \centering
  \includegraphics[width=#2]{#1}}
\newcommand {\brotatefig}[2] {\begin{figure}[htbp]
                        \centerline {
                         \epsfig{figure={#1},clip=,angle=-90,width={#2}}}}
\newcommand {\bfigfirst}[2] {\begin{figure}[h]
                        \centerline {
                        \setlength{\epsfxsize}{#2}
                        \epsffile{#1}}}
\newcommand {\efig}[2]{ \caption{#2}
                        \label{fig:#1}
                        \end{figure}
                        \mymarginpar{fig:#1}}
\newcommand {\erotatefig}[2]{ \caption{#2}
                        \label{fig:#1}
                        \end{figure}
                        \mymarginpar{fig:#1}}
\newcommand {\rfig}[1]{Figure \ref{fig:#1}}

\newcommand {\btab}[1]{
                       \begin{table}
                       \centering
                       \begin{tabular}{#1}}
\newcommand {\etab}[3] {
                       \end{tabular}
                       \caption[#3]{#2}
                       \label{tab:#1}
                       \end{table}
                       \mymarginpar{tab:#1}
                       \vspace{.1in}}

\newcommand {\btabular}[1]{\begin{center}
                       \begin{tabular}{#1}}
\newcommand {\etabular}{\end{tabular}
                       \end{center}}

\newcommand {\bdefin}[1]{\begin{definition}
                      \mymarginpar{def:#1}
                      \label{def:#1} }
\newcommand {\edefin}       {\end{definition}}
\newcommand {\rdef}[1]{Definition \ref{def:#1}}

\newcommand {\bpro}[1]{\begin{property}
                      \mymarginpar{pro:#1}
                      \label{pro:#1} }
\newcommand {\epro}   {\end{property}}

\newcommand {\bprop}[1]{\begin{proposition}
                      \mymarginpar{prop:#1}
                      \label{prop:#1} }
\newcommand {\eprop}       {\end{proposition}}
\newcommand {\rprop}[1]{Proposition \ref{prop:#1}}

\newcommand {\blem}[1]{\begin{lemma}
                      \mymarginpar{lem:#1}
                      \label{lem:#1} }
\newcommand {\elem}   {\end{lemma}}

\newcommand {\bthe}[1]{\begin{theorem}
                      \mymarginpar{the:#1}
                      \label{the:#1} }
\newcommand {\ethe}   {\end{theorem}}
\newcommand {\rthe}[1]{Theorem \ref{the:#1}}

\newcommand {\bproof}{\noindent {\bf Proof.} \ }
\newcommand {\eproof} {\hfill \squares \\ \vspace{.3cm}}
\newcommand {\bcor}[1]{\begin{corollary}
                      \mymarginpar{cor:#1}
                      \label{cor:#1} }
\newcommand {\ecor}   {\end{corollary}}
\newcommand {\rcor}[1]{Corollary \ref{cor:#1}}

\newcommand {\bax}[1]{\begin{axiom}
                      \mymarginpar{ax:#1}
                      \label{ax:#1} }
\newcommand {\eax}       {\vspace{-.1in} \end{axiom}}

\newcommand {\bex}[2]{\vspace{.1in}
                      \begin{example}
                      \mymarginpar{ex:#1}
                       {\bf #2}
                      \label{ex:#1} }
\newcommand {\eex}       {\end{example} \vspace{.3cm} }
\newcommand {\rex}[1]{Example \ref{ex:#1}}

\newcommand {\brem}[1]{\begin{remark}
                      \mymarginpar{rem:#1}
                      \label{rem:#1} \em }
\newcommand {\erem}   {\end{remark}}

\newcommand {\beq}[1]{\mymarginpar{eq:#1}
                      \begin{equation}
                      \label{eq:#1} }

\newcommand {\beqno}[1]{\mymarginpar{eq:#1}
                      \begin{eqnarray}
                      \nonumber}

\newcommand {\eeq}       {\end{equation}}
\newcommand {\eeqno}       { && \end{eqnarray}}
\newcommand {\req}[1]{(\ref{eq:#1})}

\newcommand {\bear}[1]{\mymarginpar{eq:#1}
                       \begin{eqnarray}
                       \label{eq:#1} }

\newcommand {\bearno}[1]{\mymarginpar{eq:#1}
                       \begin{eqnarray}
                       \nonumber}

\newcommand {\eear}{\end{eqnarray}}
\newcommand {\eearno}{\end{eqnarray}}
\newcommand {\bsel}{\left \{ \begin{array}{cl}}
\newcommand {\esel}{\end{array} \right.}

\newcommand {\bmat}[1]{\left [ \begin{array}{#1}}
\newcommand {\emat}{\end{array} \right ]}
\newcommand {\bsec}[2]{\mymarginpar{sec:#2}
                       \section{#1}
                       \label{sec:#2} }

\newcommand {\rsec}[1]{Section \ref{sec:#1}}

\newcommand {\bsubsec}[2]{\mymarginpar{sec:#2}
                       \subsection{#1}
                       \label{sec:#2} }

\newcommand {\rsubsec}[1]{Section \ref{sec:#1}}

\def\R{I\kern-0.30em R}
\def\N{I\kern-0.30em N}
\def\P{I\kern-0.30em P}
\newcommand\squares{\vrule height6pt width7pt depth1pt}

\newcommand{\peri}{p}

\newcommand{\cg}{\cellcolor{green}}

\begin{document}

\title{Constructions and Comparisons of Pooling Matrices for Pooled Testing of COVID-19}

\author{Yi-Jheng Lin,  Che-Hao Yu, Tzu-Hsuan Liu, Cheng-Shang~Chang,~\IEEEmembership{Fellow,~IEEE,}\\ and Wen-Tsuen Chen,~\IEEEmembership{Life Fellow,~IEEE}
                \thanks{Y.-J. Lin, C.-H. Yu, T.-H. Liu, C.-S. Chang, and W.-T. Chen are with the Institute of Communications Engineering, National Tsing Hua University, Hsinchu 30013, Taiwan, R.O.C. Email:  s107064901@m107.nthu.edu.tw; chehaoyu@gapp.nthu.edu.tw;  carina000314@gmail.com;  cschang@ee.nthu.edu.tw;  wtchen@cs.nthu.edu.tw.}
}

\maketitle
\begin{abstract}
	In comparison with individual testing, group testing is more efficient in reducing the number of tests and potentially leading to tremendous cost reduction. There are two key elements in a group testing technique: (i) the pooling matrix that directs samples to be pooled into groups, and (ii) the decoding algorithm that uses the group test results to reconstruct the status of each sample. In this paper, we propose a new family of pooling matrices from packing the pencil of lines (PPoL) in a finite projective plane. We compare their performance with various pooling matrices proposed in the literature, including 2D-pooling, P-BEST, and Tapestry, using the two-stage definite defectives (DD) decoding algorithm. By conducting extensive simulations for a range of prevalence rates up to 5\%, our numerical results show that there is no pooling matrix with the lowest relative cost in the whole range of the prevalence rates. To optimize the performance, one should choose the right pooling matrix, depending on the prevalence rate. The family of PPoL matrices can dynamically adjust their construction parameters according to the prevalence rates and could be a better alternative than using a fixed pooling matrix.
\end{abstract}





%

\bsec{Introduction}{introduction}

COVID-19 pandemic has deeply affected the daily life of many people in the world.
The current strategy for dealing with COVID-19 is to reduce the transmission rate of COVID-19 by preventive measures, such as
contact tracing, wearing masks, and social distancing. One problematic characteristic of COVID-19 is that there are asymptomatic infections \cite{who}. As those asymptomatic infections are unaware of their contagious ability, they can infect more people if they are not yet been detected \cite{nishiura2020estimation}.
As shown in the recent paper \cite{chen2020time}, massive COVID-19 testing in South Korea on Feb. 24, 2020, can greatly reduce the proportion of undetectable infected persons and effectively reduce the transmission rate of COVID-19.

Massive testing for a large population is very costly if it is done one at a time.
For a population with a low prevalence rate, group testing (or pool testing, pooled testing, batch testing) that tests a group by mixing several samples together can achieve a great extent of saving testing resources.
As indicated in the recent article posted on the US FDA website \cite{fdaAug}, the group testing approach has received a lot of interest lately. Also, in the US CDC's guidance for the use of pooling procedures in SARS-CoV-2 \cite{cdcAug},
it defines three types of tests: (i) {\em diagnostic testing} that is intended to identify occurrence at the individual level and is performed when there is a reason to suspect that an individual may be infected, (ii)
{\em screening testing} that is intended to identify occurrence at the individual level even if there is no reason to suspect an infection, and (iii) {\em surveillance testing} includes ongoing systematic activities, including collection, analysis, and interpretation of health-related data. The general guidance for diagnostic or screening testing using a pooling strategy in \cite{cdcAug} (quoted below) basically follows the two-stage group testing procedure invented by Dorfman in 1943 \cite{dorfman1943detection}:

{\em
	``If a pooled test result is negative, then all specimens can be presumed negative with the single test. If the test result is positive or indeterminate, then all the specimens in the pool need to be retested individually.''}


The Dorfman two-stage algorithm is a very simple group testing strategy.
Recently, there are more sophisticated group testing algorithms proposed in the literature, see, e.g., \cite{sinnott2020evaluation,shental2020efficient,ghosh2020tapestry,ghosh2020compressed}. Instead of pooling a sample into a single group, these algorithms require diluting a sample and then splitting it into multiple groups (pooled samples). Such a procedure is specified by
a {\em pooling matrix} that directs each diluted sample to be pooled into a specific group.  The test results of pooled samples are then used for
decoding (reconstructing) the status of each sample. In short, there are two key elements in a group testing strategy: (i) the pooling matrix, and (ii) the decoding algorithm.

As COVID-19 is a severe contagious disease, one should be very careful about the decoding algorithm
used for reconstructing the testing results of persons.
Though decoding algorithms that use soft information for group testing, including various  compressed sensing algorithms in \cite{shental2020efficient,yi2020low,ghosh2020tapestry,ghosh2020compressed,heidarzadeh2020two}, might be more efficient in reducing the number of tests, they are more prone to have false positives and false negatives. A false positive might cause a person to be quarantined for 14 days, and thus losing 14 days of work. On the other hand, a false negative might have an infected person wandering around  the neighborhood and cause more people to be infected.
In view of this, it is important to have group testing results that are as ``definite'' as individual testing results (in a noiseless setting).

Following the CDC guidance \cite{cdcAug},
we use
the decoding algorithm, called
the {\em definite defectives (DD)} algorithm in the literature (see Algorithm 2.3 of the monograph \cite{aldridge2019group}),  that can have definite testing results.
The DD algorithm first identifies negative samples from a negative testing result of a group (as advised by the CDC guidance \cite{cdcAug}). Such a step is known as the combinatorial orthogonal matching pursuit (COMP) step in the literature \cite{aldridge2019group}. Then the DD algorithm identifies positive samples if they are in a group with only one positive sample. Not every sample can be decoded by the DD algorithm.
As the Dorfman two-stage algorithm, samples that are not decoded by the DD algorithm go through the second stage, and they are tested individually. We call such an algorithm the two-stage DD algorithm.

One of the main objectives of this paper is to compare the performance of various pooling matrices proposed in the literature, including 2D-pooling \cite{sinnott2020evaluation}, P-BEST \cite{shental2020efficient}, and Tapestry \cite{ghosh2020tapestry,ghosh2020compressed}, using the two-stage DD decoding algorithm. In addition to these pooling matrices, we also propose a new construction of a family of pooling matrices from packing the pencil of lines (PPoL) in a finite projective plane.
The family of PPoL pooling matrices has very nice properties: (i) both the column correlation and the row correlation are bounded by 1, and (ii) there is a freedom to choose the construction parameters to optimize performance.
To measure the amount of saving of a group testing method, we adopt the performance measure, called the {\em expected relative cost} in \cite{dorfman1943detection}. The expected relative cost is defined as the ratio of the expected number of tests required by the group testing technique to the number of tests required by the individual testing.
We then measure the expected relative costs of these pooling matrices  for a range of prevalence rates up to 5\%.
Some of the main findings of our numerical results are as follows:
\begin{description}
	\item[(i)] There is no pooling matrix that has  the lowest relative cost in the whole range of the prevalence rates considered in our experiments. To optimize the performance, one should choose the right pooling matrix, depending on the prevalence rate.
	
	\item[(ii)] The expected relative costs of the two pooling matrices used in Tapestry \cite{ghosh2020tapestry,ghosh2020compressed}
	are high compared to the other pooling matrices considered in our experiments. Its performance, in terms of the expected relative cost, is even worse than the (optimized) Dorfman two-stage algorithm.
	However, Tapestry is capable of decoding most of the samples in the first stage. In other words, the percentages of samples that need to go through the second stage are the smallest among all the pooling matrices considered in our experiments.
	
	\item[(iii)] P-BEST \cite{shental2020efficient} has a very low expected relative cost when the prevalence rate is below 1\%. However, its expected relative cost increases dramatically when the prevalence rate is above 1.3\%.

	\item[(iv)] 2D-pooling \cite{sinnott2020evaluation}  has a low expected relative cost when the prevalence rate is near 5\%. Unlike Tapestry, P-BEST, and PPoL that rely on robots for pipetting, the implementation of 2D-pooling is relatively easy by humans.

	\item[(v)] There is a PPoL pooling matrix with column weight 3 that outperforms the P-BEST pooling matrix for the whole range of the prevalence rates considered in our experiments (up to 5\%).
	We suggest using that PPoL pooling matrix up to the prevalence rate of 2\% and then switch to other PPoL pooling matrices with respect to the increase of the prevalence rate. The detailed suggestions are shown in Table \ref{tab:suboptimal} of \rsec{num}.
\end{description}

The paper is organized as follows: in \rsec{binary}, we briefly review the group testing problem, including the mathematical formulation and the DD decoding algorithm. In \rsec{works}, we introduce the related works that are used in our comparison study. We then propose the new family of PPoL pooling matrices in \rsec{constructionsample}.
In \rsec{num}, we conduct extensive simulations to compare the performance of various pooling matrices using the two-stage DD algorithm. The paper is concluded in \rsec{con}, where we discuss possible extensions for future works.

\bsec{Review of Group Testing}{binary}

\bsubsec{The problem statement}{problem}

\begin{figure}
	\centering
	\includegraphics[width=0.30\textwidth]{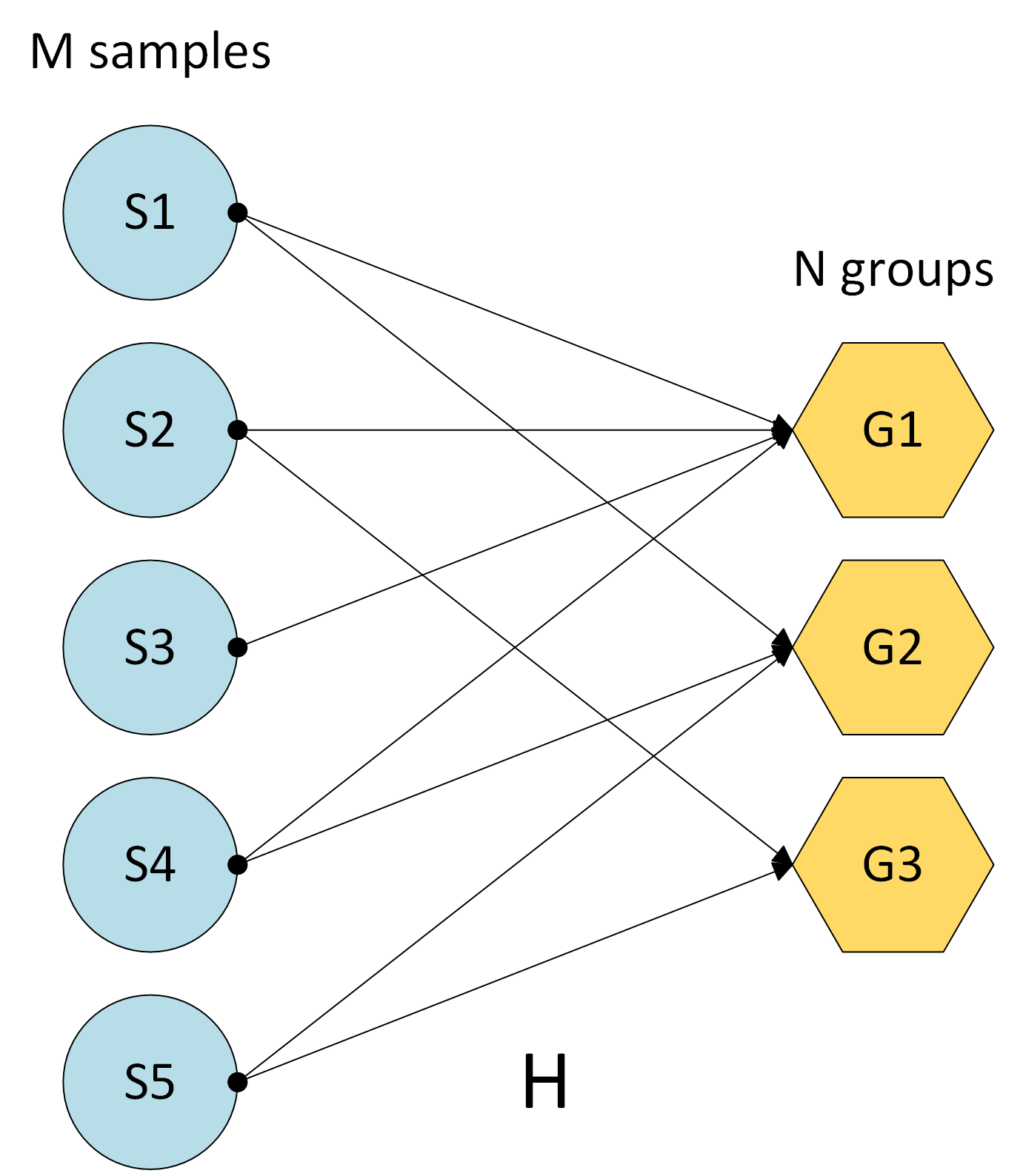}
	\caption{Pooled testing represented by a bipartite graph.}
	\label{fig:pooled}
\end{figure}

Consider the group testing problem with $M$ samples (indexed from $1,2, \ldots, M$), and $N$ groups (indexed from $1,2, \ldots, N$).
The $M$ samples are pooled into the $N$ groups (pooled samples) through an $N \times M$ binary matrix $H=(h_{n,m})$ so that
the $m^{th}$ sample is pooled into the $n^{th}$ group if $h_{n,m}=1$ (see \rfig{pooled}). Such a matrix is called the {\em pooling matrix} in this paper.
Note that a pooling matrix corresponds to the biadjacency matrix of an $N \times M$ bipartite graph.
Let $x=(x_1, x_2, \ldots, x_M)$ be the binary state vector of the $M$ samples and
$y=(y_1, y_2, \ldots, y_N)$ be the binary state vector of the $N$ groups.
Then
\beq{groupt1111}
y=Hx,
\eeq
where the matrix operation is under the Boolean algebra (that replaces the usual addition by the OR operation and the usual multiplication by the AND operation).
The main objective of group testing is to decode the vector $x$ given the observation vector $y$ under certain assumptions.
In this paper, we adopt the following basic assumptions for binary samples:
\begin{description}
	\item[(i)] Every sample is binary, i.e., it is either positive (1) or negative (0).
	\item[(ii)] Every group is binary, and a group is positive (1) if there is at least one sample in that group is positive. On the other hand, a group is negative (0) if all the samples pooled into that group are negative.
\end{description}

If we test each sample one-at-a-time, then the number of tests for $M$ samples is $M$, and the average number of tests per sample is 1.
The key advantage of using group testing is that the number of tests per sample can be greatly reduced.
One important performance measure of group testing, called the {\em expected relative cost} in \cite{dorfman1943detection}, is the ratio of the expected number of tests required by the group testing technique to the number of tests required by the individual testing.
The main objective of this paper is to compare the expected relative costs of various group testing methods.

\bsubsec{The definite defectives (DD) decoding algorithm}{decoding}

In this section, we briefly review
the definite defectives (DD) algorithm (see Algorithm 2.3 of \cite{aldridge2019group}).
The DD algorithm first identifies negative samples from a negative testing result of a group. Such a step is known as the combinatorial orthogonal matching pursuit (COMP) step. Then the DD algorithm identifies positive samples if they are in a group with only one positive sample.
The detailed steps of the DD algorithm are outlined in Algorithm \ref{alg:binary}.

\begin{algorithm}\caption{The definite defectives (DD) algorithm  for binary samples}\label{alg:binary}
	
	\noindent {\bf Input}  An $N\times M$ pooling matrix $H$ and a binary $N$-vector $y$ of the group test result.
	
	\noindent {\bf Output} an $M$-vector for the test results of the $M$ samples.
	
	\noindent 0: Initially, every sample is marked ``un-decoded.''
	
	\noindent 1: If there is a negative group, then all the samples pooled into that group are decoded to be negative.
	
	\noindent 2: The edges of samples decoded to be negative in the bipartite graph are removed from the graph.
	
	\noindent 3: Repeat from Step 1 until there is no negative group.
	
	\noindent 4: If there is a positive group with exactly one (remaining) sample in that group, then that sample is decoded to positive.
	
	\noindent 5:  Repeat from Step 4 until no more samples can be decoded.

\end{algorithm}

\begin{figure*}[tb]
	\begin{center}
		\begin{tabular}{p{0.3\textwidth}p{0.3\textwidth}p{0.3\textwidth}}
			\includegraphics[width=0.3\textwidth]{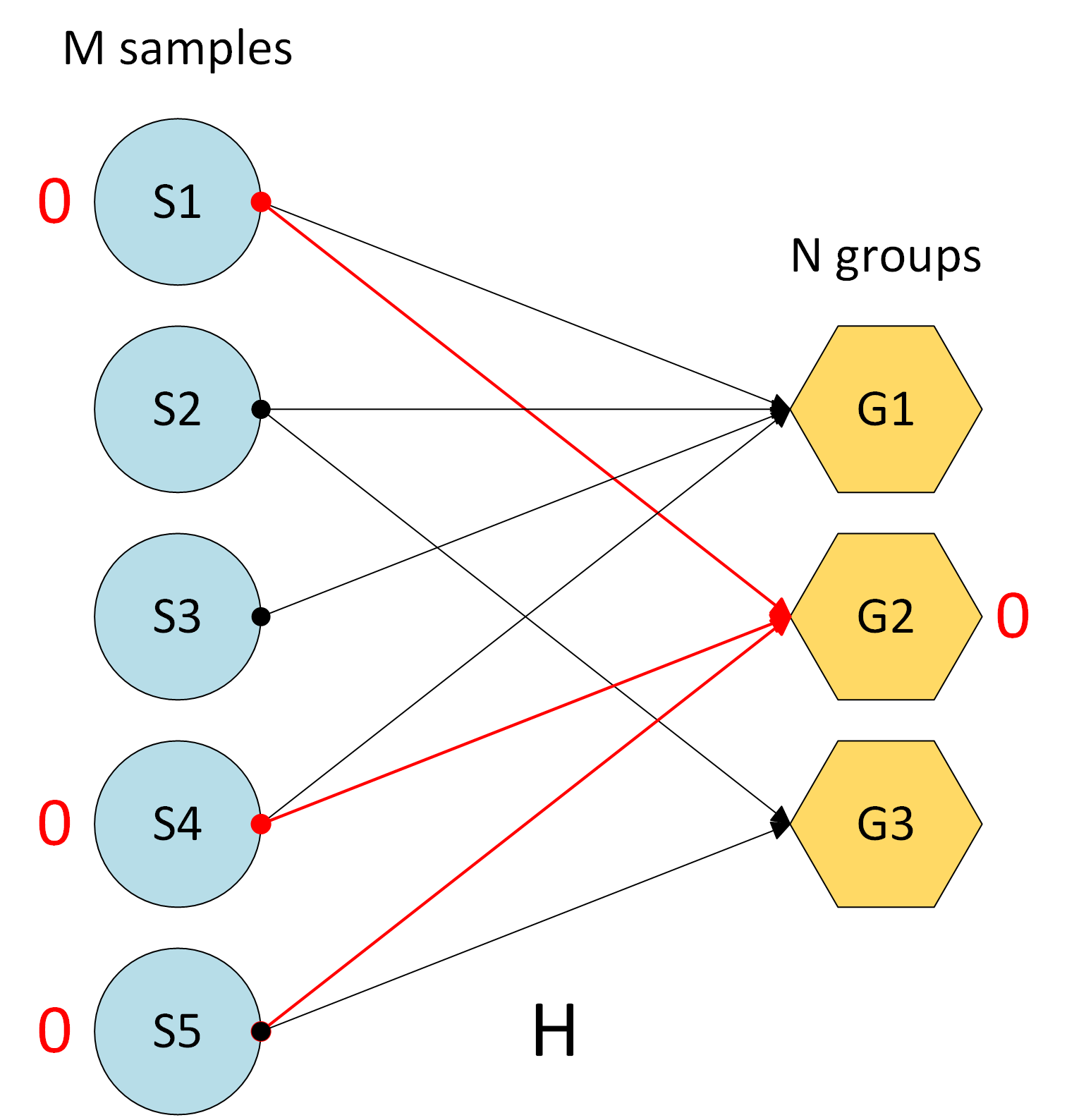} &
			\includegraphics[width=0.3\textwidth]{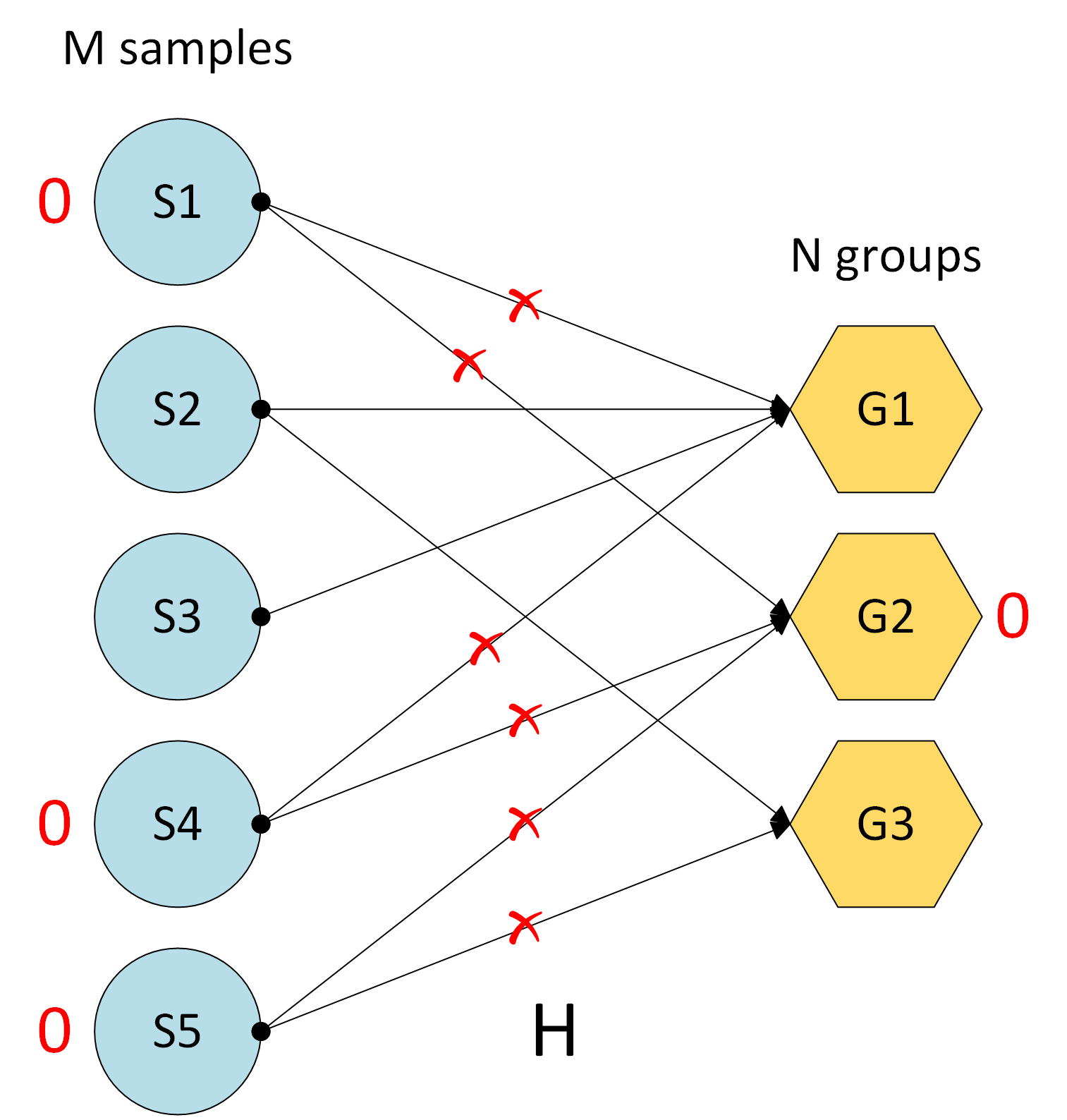} &
			\includegraphics[width=0.3\textwidth]{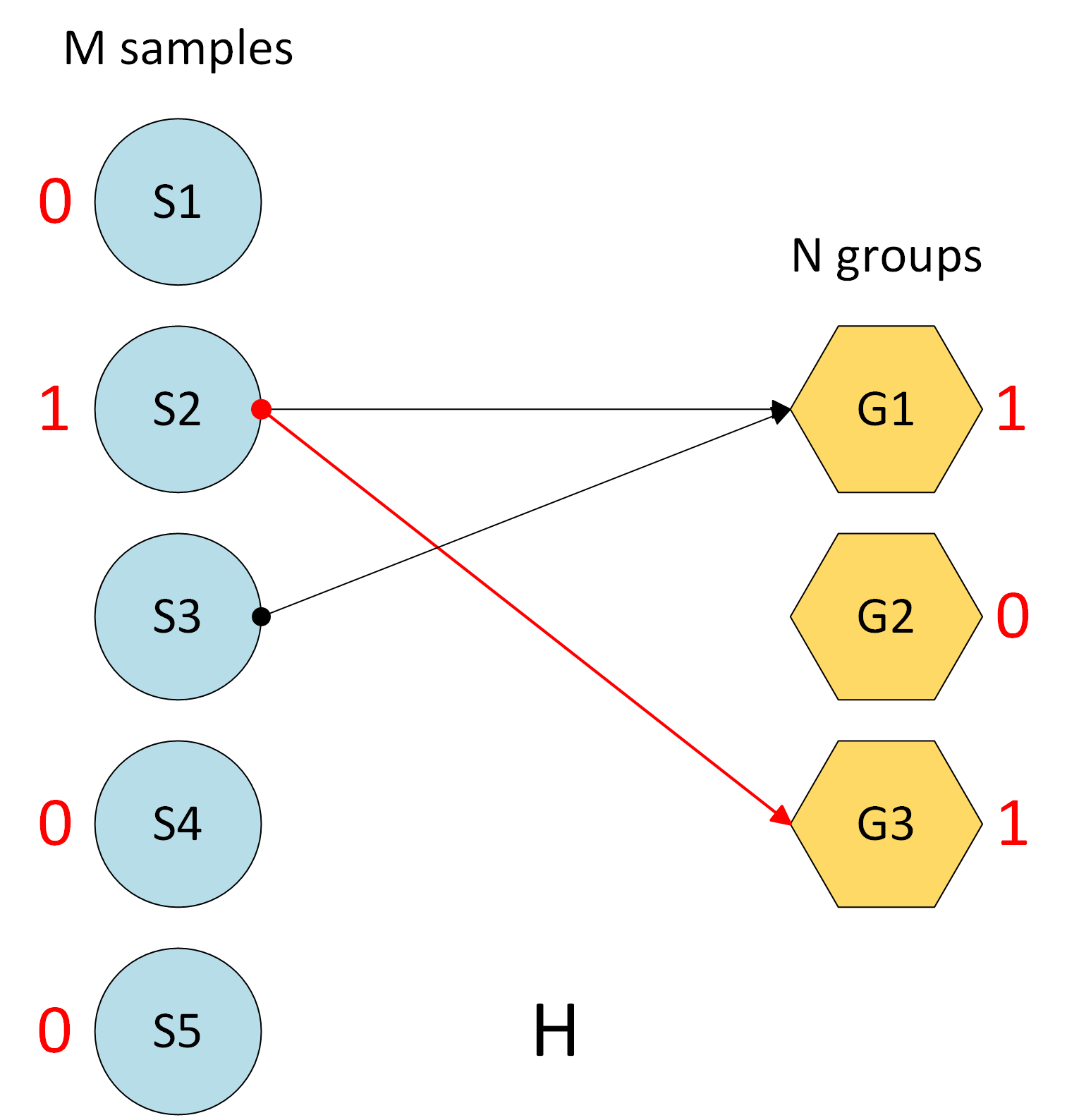}\\
			(a) Step 1: All the samples pooled into that negative groups are decoded to be negative.
			& (b) Step 2: The edges  of negative samples are removed. & (c) Step 4: Exactly one sample in a positive group is decoded to be positive. \\
		\end{tabular}
		\caption{An illustration for the DD algorithm.}
		\label{fig:decoderb}
	\end{center}
\end{figure*}

In \rfig{decoderb}, we provide an illustrating example for Algorithm \ref{alg:binary}. In \rfig{decoderb} (a), the test result of $G2$ is negative, and thus the three samples $S1$, $S4$ and $S5$, are decoded to be {\em negative}. In \rfig{decoderb} (b), the edges that are connected to the
samples $S1$, $S4$ and $S5$, are removed from the bipartite graph. In \rfig{decoderb} (c), the test results of the two groups $G1$ and $G3$ are positive. As $S2$ is the only sample in $G3$, $S2$ is decoded to be {\em positive}.

\begin{figure}
	\centering
	\includegraphics[width=0.30\textwidth]{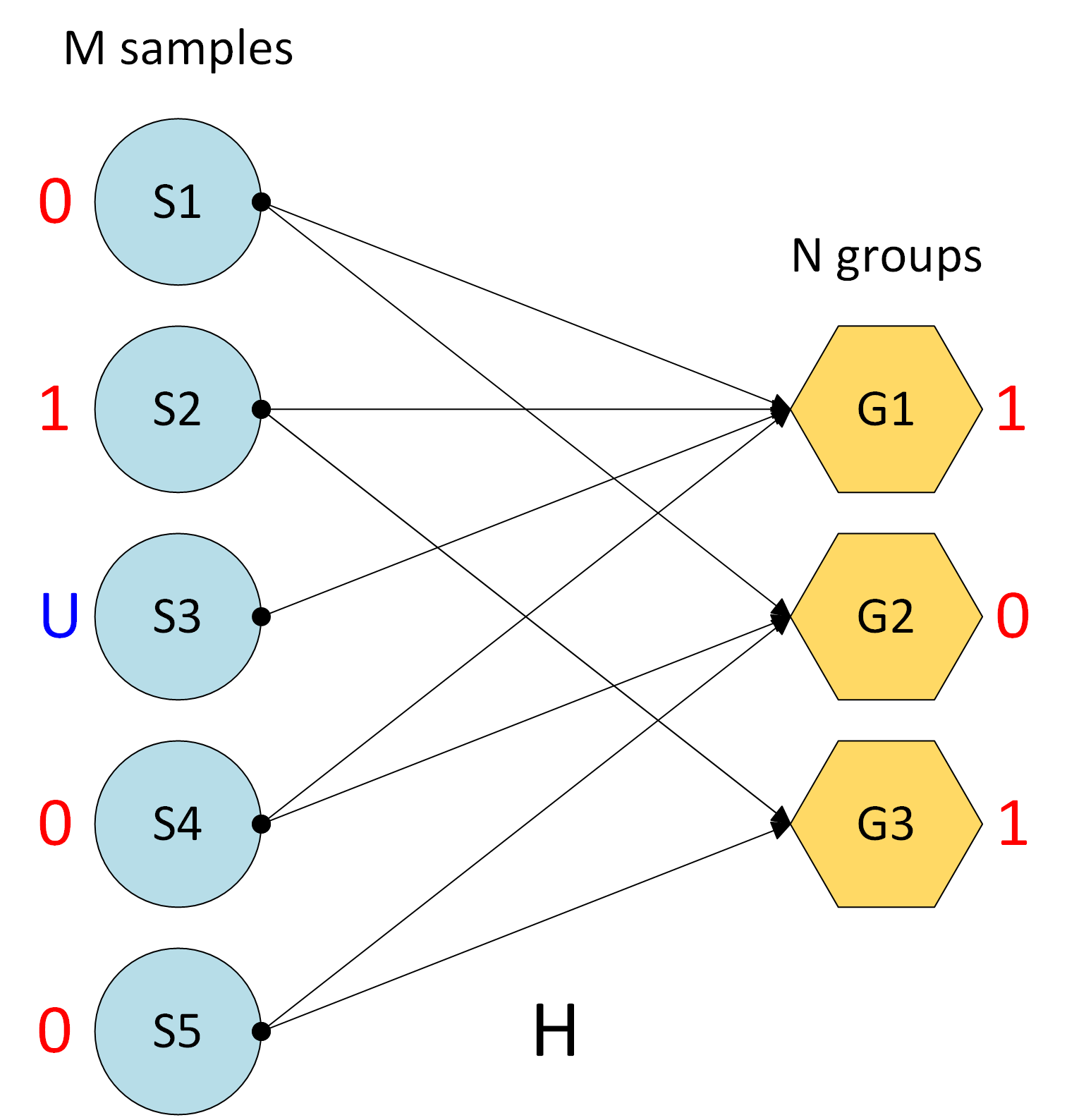}
	\caption{An un-decoded sample.}
	\label{fig:undecode}
\end{figure}

Note that one might not be able to decode all the samples by the above decoding algorithm. For instance, if a particular sample is pooled into groups that all have at least one positive sample, then there is no way to know whether that sample is positive or negative.
As shown in \rfig{undecode}, the sample $S3$ cannot be decoded by the DD algorithm as the test results of the three groups are the same no matter if $S3$ is positive or not.

As shown in Lemma 2.2 of \cite{aldridge2019group}, one important guarantee of the DD algorithm is that
there is no false positive.

\bprop{DD1}(\cite{aldridge2019group}, Lemma 2.2)
Assume that all the testing results are correct. Then (i) all the samples that are decoded to be negative in Step 1 of Algorithm \ref{alg:binary}
are definite negatives, and (ii) all the samples that are decoded to be positive in Step 4 of Algorithm \ref{alg:binary}
are definite positives. As such,
there are no false positives in Algorithm \ref{alg:binary}.
\eprop

In order to resolve all the ``un-decoded'' samples, we add another stage by individually testing each ``un-decoded'' sample.
This leads to the following two-stage DD algorithm in Algorithm \ref{alg:binary2}.

\begin{algorithm}\caption{The two-stage definite defectives (DD2) algorithm  for binary samples}\label{alg:binary2}
	
	\noindent {\bf Input}  An $N\times M$ pooling matrix $H$ and a binary $N$-vector $y$ of the group test result.
	
	\noindent {\bf Output} an $M$-vector for the test results of the $M$ samples.
	
	
	\noindent 1: Run the DD algorithm in Algorithm \ref{alg:binary}.
	
	\noindent 2:  For those ``un-decoded'' samples, test them one at a time.
	
\end{algorithm}

\bsec{Related Works}{works}

In \cite{lohse2020pooling,abdalhamid2020assessment,yelin2020evaluation}, it was shown that a single positive sample can still be
detected even in pools of 5-32 samples
for the standard RT-qPCR test of COVID-19. Such an experimental result provides supporting evidence for group testing of COVID-19.
In the following, we review four group testing strategies proposed in the literature for COVID-19.

\noindent {\bf The Dorfman two-stage algorithm \cite{gollier2020group}}: For the case that $N=1$, i.e., every sample is pooled into a single group, the DD2 algorithm is simply the original Dorfman two-stage algorithm \cite{dorfman1943detection}, i.e., if the group of $M$ samples is tested negative, then all the $M$ samples are ruled out. Otherwise, all the $M$ samples are tested individually.
Suppose that the prevalence rate is $r_1$. Then the expected number of tests to decode the $M$ samples by the Dorfman two-stage algorithm is
$1+(1-(1-r_1)^M)M$. As such, the expected relative cost (defined as the ratio of the expected number of tests required by the group testing technique to the number of tests required by the individual testing in \cite{dorfman1943detection}) is $\frac{M+1}{M}-(1-r_1)^M$.
As shown in Table I of \cite{dorfman1943detection}, the optimal group size $M$ is 11 with the expected relative cost of 20\% when the prevalence rate $r_1$ is 1\%.

\noindent{\bf 2D-pooling \cite{sinnott2020evaluation}}: On a 96-well plate, there are 8 rows and 12 columns.
Pool the samples in the same row (column) into a group. This results in 20 groups for 96 samples. One advantage of this simple 2D-pooling strategy
is to minimize pipetting errors.

\noindent{\bf P-BEST \cite{shental2020efficient}}: P-BEST \cite{shental2020efficient} uses a $48 \times 384$ pooling matrix constructed from the Reed-Solomon code \cite{reed1960polynomial} for pooled testing of COVID-19. For the pooling matrix, each sample is pooled into 6 groups, and each group contains 48 samples.
In \cite{shental2020efficient}, the authors proposed using a compressed sensing algorithm, called the Gradient Projection for Sparse Reconstruction (GPSR) algorithm for decoding. 
Though it is claimed in \cite{shental2020efficient} that the GPSR algorithm can detect up to 1\% of positive carriers, there is no guarantee that every decoded sample (by the GPSR algorithm) is correct.

\noindent {\bf Tapestry \cite{ghosh2020tapestry,ghosh2020compressed}}: The Tapestry scheme \cite{ghosh2020compressed,ghosh2020tapestry} uses the Kirkman triples to construct their pooling matrices. For the pooling matrix in \cite{ghosh2020compressed,ghosh2020tapestry}, each sample is pooled into 3 groups (in their experiments, some samples are only pooled into 2 groups). 
As such, it is sparser than that used by P-BEST. However, one of the restrictions for the pooling matrices constructed from the Kirkman triples is that the column weights must be 3. Such a restriction limits its applicability to optimize its performance according to the prevalence rate.
We note that a compressed-sensing-based decoding algorithm was proposed in \cite{ghosh2020tapestry,ghosh2020compressed}. Such a decoding algorithm further exploits the viral load (Ct value) of each pool and reconstructs the Ct value of each positive sample.
It is claimed to be viable not just with low ($<4$\%) prevalence rates, but even with moderate prevalence rates (5\%-10\%). 


	
\bsec{PPoL Constructions of Pooling Matrices}{constructionsample}
	
In this section, we propose a new family of pooling matrices from packing the pencil of lines (PPoL) in
a finite projective plane. Our idea of constructing PPoL pooling matrices was inspired by the constructions of channel hopping sequences in the rendezvous search problem in cognitive radio networks and the constructions of grant-free uplink transmission schedules in 5G networks (see, e.g., \cite{lin2021ppol,chang2015multichannel,chang2019asynchronous,chang2019theoretical}), in particular, the channel hopping sequences constructed by the PPoL algorithm in \cite{lin2021ppol}.
	
A pooling matrix is said to be $(d_1,d_2)$-regular if there are exactly $d_1$ (resp. $d_2$) nonzero elements in each column (resp. row).
In other words, the degree of every left-hand (resp. right-hand) node in the corresponding bipartite graph  is $d_1$ (resp. $d_2$).
The total number of edges in the bipartite graph is $d_1 M=d_2 N$ for a  $(d_1,d_2)$-regular pooling matrix $H$.
Define the (compressing) gain
\beq{groupt2222}
	G=\frac{M}{N}=\frac{d_2}{d_1}.
\eeq

\bsubsec{Perfect difference sets and finite projective planes}{fpp}
	
As our construction of the pooling matrix is from packing the pencil of lines in a finite projective plane,
we first briefly
review the notions of difference sets and  finite projective planes.
	
\bdefin{difference}{\bf (Difference sets)}
Let $Z_\peri=\{0,1,\ldots, \peri-1\}$.
A set $D = \{a_0, a_1, \ldots , a_{k-1}\} \subset Z_\peri$  is called a $(\peri,k,\lambda)$-difference set if for every $(\ell\;\mbox{mod}\;\peri) \ne 0$, there exist
at least $\lambda$ ordered pairs $(a_i, a_j)$ such that $a_i-a_j = (\ell\;{\rm mod}\;\peri)$, where $a_i, a_j\in  D$.
A $(\peri,k,1)$-difference set is said to be {\em perfect} if  there exists
exactly one  ordered pair $(a_i, a_j)$ such that $a_i-a_j = (\ell\;{\rm mod}\;\peri)$ for every $(\ell\;\mbox{mod}\;\peri) \ne 0$.
\edefin

\bdefin{fpp}{\bf (Finite projective planes)}
	A finite projective plane of order $m$, denoted by $PG(2,m)$, is a collection of $m^2 +m+1$ lines and $m^2+m+1$ points such that
	\begin{description}
		\item[(P1)] every line contains $m+1$ points,
		\item[(P2)] every point is on $m+1$ lines,
		\item[(P3)] any two distinct lines intersect at exactly one point, and
		\item[(P4)] any two distinct points lie on exactly one line.
	\end{description}
\edefin
	
When $m$ is a prime power, Singer \cite{Singer1938} established the connection between an $(m^2+m+1, m+1,1)$-perfect difference set and a finite projective plane of order $m$ through a collineation that maps points (resp. lines) to points (resp. lines) in a finite projective plane.
Specifically, suppose that $D = \{a_0, a_1, \ldots , a_{m}\}$ is an  $(m^2+m+1, m+1,1)$-perfect difference set
with
\beq{diff1111a}
	a_0=0 <a_1=1 < a_2 < \ldots, < a_{m} < m^2+m+1.
\eeq

\begin{description}
	\item[(i)] Let
	$\{0,1,\ldots, m^2+m\}$ be the $m^2+m+1$ points.
	\item[(ii)] Let  $\peri=m^2+m+1$ and $D_\ell=\{(a_0+\ell)\;\mbox{mod}\;\peri, (a_1+\ell)\;\mbox{mod}\;\peri, \ldots , (a_{m}+\ell)\;\mbox{mod}\;\peri\}$, $\ell=0,1,2,\ldots, \peri-1$
	be the $m^2+m+1$ lines.
\end{description}
Then these $m^2+m+1$ points and $m^2+m+1$ lines form a finite projective plane of order $m$.

\bsubsec{The construction algorithm}{algorithm}

In this section, we propose the PPoL algorithm for constructing pooling matrices.
For this, one first constructs an $(m^2+m+1, m+1,1)$-perfect difference set, $D = \{a_0, a_1, \ldots , a_{m}\}$ with
\beq{diff1111c}
a_0=0 <a_1=1 < a_2 < \ldots, < a_{m} < m^2+m+1.
\eeq
Let $\peri=m^2+m+1$ and
\beq{trans1157}
D_\ell=\{(a_0+\ell)\;{\rm mod}\;\peri, (a_1+\ell)\;{\rm mod}\;\peri, \ldots , (a_{m}+\ell)\;{\rm mod}\;\peri\},\eeq
$\ell=0,1,2,\ldots, \peri-1$
be the $\peri$ lines in the corresponding finite projective plane.

It is easy to see that the $m+1$ lines in the corresponding finite projective plane that contain point $0$ are $D_0, D_{\peri-a_1}, D_{\peri-a_2}, \ldots, D_{\peri-a_m}$.
These $m+1$ lines are called the pencil of lines that contain point 0 (as the pencil point). As the only intersection of the $m+1$ lines is point 0,
these $m+1$ lines, excluding point 0, are  disjoint, and thus
can be packed into $Z_\peri$. This is formally proved in the following lemma.

\blem{disjoint}
Let $D_{\peri-a_i}^0 =D_{\peri-a_i}\backslash \{0\}$, $i=1,2, \ldots, m$.
Then  $\{D_0, D_{\peri-a_1}^0, \ldots, D_{\peri-a_m}^0\}$ is a partition of $Z_\peri$.
\elem

\bproof
First, note that
$\{D_0, D_{\peri-a_1}, \ldots, D_{\peri-a_m}\}$ are the $m+1$ lines that contain point 0.
As any two distinct lines intersect at exactly one point,
we know that for $i \ne 0$,
$$D_0 \cap D_{\peri-a_i}^0= \varnothing,$$
and that for $i \ne j$,
$$ D_{\peri-a_i}^0 \cap D_{\peri-a_j}^0= \varnothing .$$
Thus, they are disjoint.

As there are $m+1$ points in $D_0$ and $m$ points in $D_{\peri-a_i}^0$,
$D_0 \cup D_{\peri-a_1}^0\cup \ldots \cup D_{\peri-a_m}^0$ contains $m+1 + m^2$ points.
These $m+1 + m^2$ points are exactly the set of $m^2+m+1$ points in the finite projective plane of order $m$.
\eproof

In Algorithm \ref{alg:PPoLs}, we show how one can construct a pooling matrix from a finite projective plane.
The idea is to first construct a bipartite graph with the line nodes on the left and the point nodes  on the right.
There is an edge between a point node and a line  node if that point is in that line.
Then we start trimming this line-point bipartite graph to achieve the needed compression ratio.
Specifically, we select the subgraph with the $m^2$ line nodes that does not contain point 0 (on the left) and
the $d_1 m$ point nodes in the union of $d_1$ pencil of lines (on the right).

Note that in Algorithm \ref{alg:PPoLs}, the number of samples has to be $m^2$. However, this restriction may not be met in practice. A simple way to tackle this problem is by adding additional dummy samples to ensure that the total number of samples is $m^2$.
In the literature, there are some sophisticated methods (see, e.g., the recent work \cite{hong2021hyper}) that further consider the ``balance'' issue, i.e., samples should be pooled into groups as evenly as possible.

\begin{algorithm}\caption{The PPoL algorithm}\label{alg:PPoLs}
	
	\noindent {\bf Input}  The number of samples $M=m^2$ with $m$ being a prime power, and the degree of each sample $1 \le d_1 \le m+1$.
	
	\noindent {\bf Output} An $N \times M$ binary pooling matrix $H$ with $M=m^2$ and $N=d_1 m$.
	
	\noindent 1: Let $\peri=m^2+m+1$ and construct a perfect difference set $D = \{a_0, a_1, \ldots , a_{m}\}$ in $Z_\peri$ (with $a_0=0$ and $a_1=1$).
	
	\noindent 2: For $\ell=0,1, \ldots ,\peri-1$, let
	$$D_\ell=\{(a_0+\ell)\;{\rm mod}\;\peri, (a_1+\ell)\;{\rm mod}\;\peri, \ldots , (a_{m}+\ell)\;{\rm mod}\;\peri\}$$
	be the $\peri$ lines.
	
	\noindent 3: Construct a bipartite graph with the $\peri$ lines on the left and the $\peri$ points on the right.
	Add an edge between a point node and a line node if that point is in that line.
	
	\noindent 4: Remove point 0 and line 0 from the bipartite graph (and the edges attached to these two nodes). Let $G=(g_{n,\ell})$ be the $(m^2+m) \times (m^2+m)$ biadjacency matrix
	of the trimmed bipartite graph with $g_{n,\ell}=1$ if point $n$ is in $D_\ell$.
	
	\noindent 5: Let $D_{\peri-a_i}^0 =D_{\peri-a_i}\backslash \{0\}$, $i=0,1,2, \ldots, m$, be the $m+1$ pencil of lines that contain point 0.

	\noindent 6: Remove the ${(\peri-a_i)}^{th}$ column, $i=1,2, \ldots, m$, in $G$ to form an $(m^2+m) \times m^2$ biadjacency matrix $\tilde G$.
	Note that these $m$ columns correspond to the $m$ lines containing point 0.
	
	\noindent 7: Let $B=\cup_{i=0}^{d_1-1} D_{\peri-a_i}^0$ (select the first $d_1$ pencil of lines that contain point 0). Remove rows of $\tilde G$ that are not in $B$ to form a
	$d_1 m \times m^2$ biadjacency matrix $H$.
\end{algorithm}

\begin{figure*}[tb]
	\begin{center}
		\begin{tabular}{p{0.3\textwidth}p{0.3\textwidth}p{0.3\textwidth}}
			\includegraphics[width=0.3\textwidth]{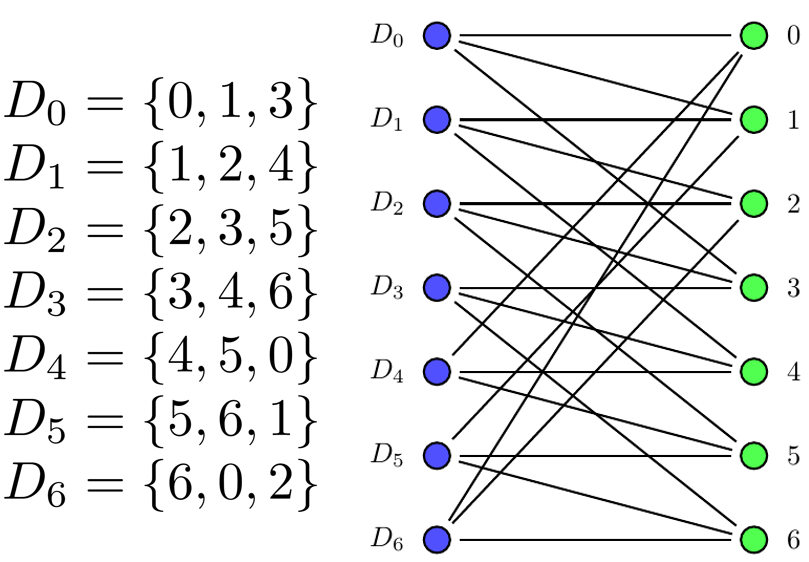} &
			\includegraphics[width=0.3\textwidth]{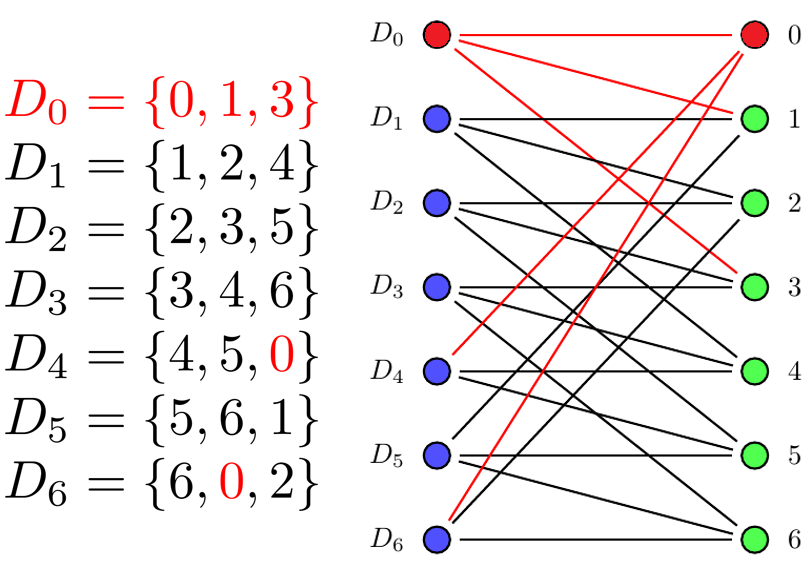} &
			\includegraphics[width=0.3\textwidth]{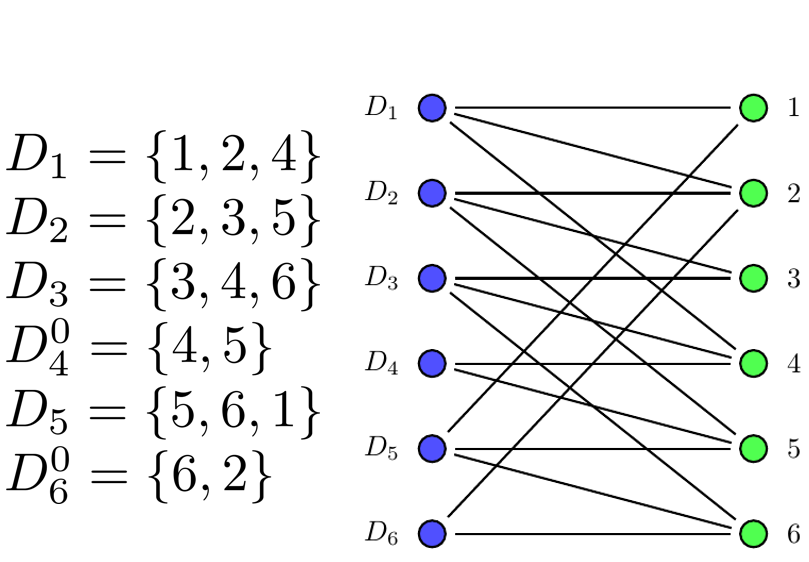}\\
			  (a) Step 3: The bipartite graph with $7$ lines and $7$ points.
			& (b) Step 4: The nodes and the edges that need to be removed are marked in red. 
			& (c) Step 4: The trimmed bipartite graph. \\
			\includegraphics[width=0.3\textwidth]{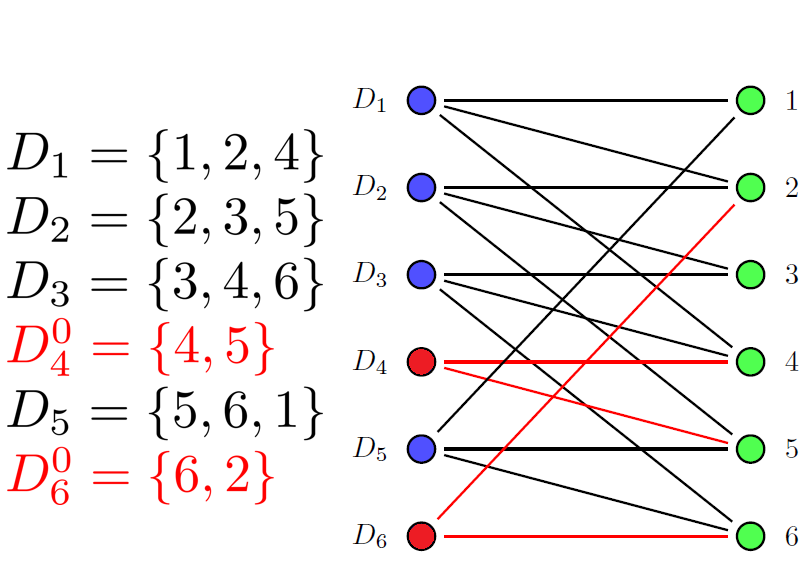} &
			\includegraphics[width=0.3\textwidth]{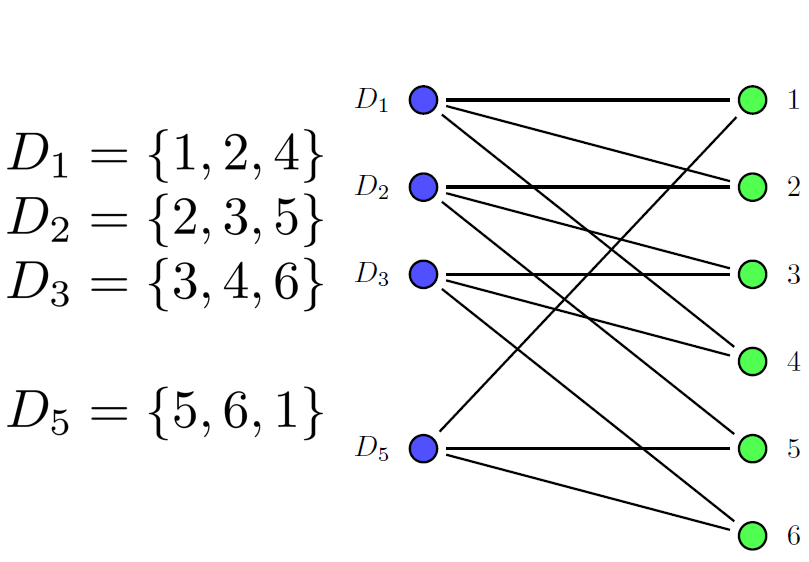} &
			\includegraphics[width=0.3\textwidth]{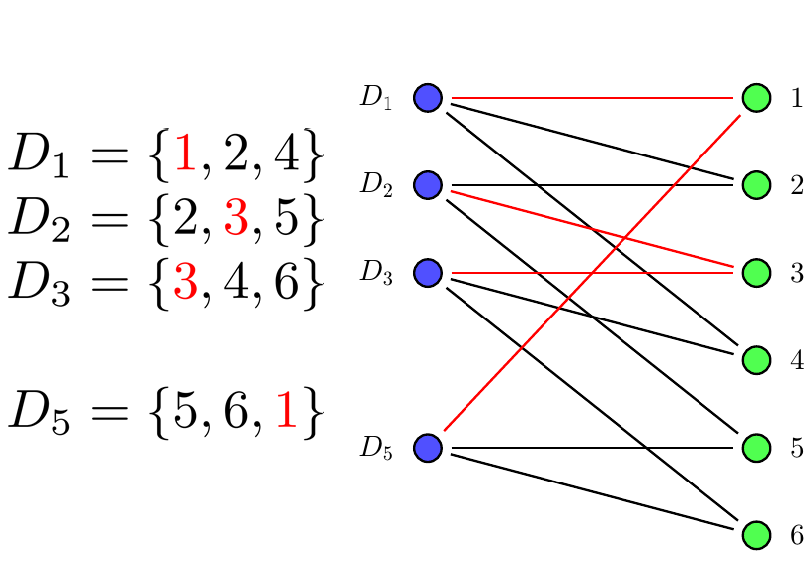}\\
			  (d) Step 6: The two lines that need to be removed are marked in red.
			& (e) Step 6: The bipartite graph after removing the two lines.
			& (f) Step 7: The points in set $B$ along with the edges attached to these nodes are marked in red.
		\end{tabular}
		\caption{An example to demonstrate how the PPoL algorithm in Algorithm \ref{alg:PPoLs} works.}
		\label{fig:algoex}
	\end{center}
\end{figure*}

{\color{black}
	\bex{algoex}{(A worked example of the PPoL algorithm in Algorithm \ref{alg:PPoLs})}
	Let $m=2$, $d_1=1$ be the inputs of Algorithm \ref{alg:PPoLs}. 
	In Step 1, let $p=m^2+m+1=7$ and construct the perfect difference set $D=\{a_0,a_1,a_2\} = \{0,1,3\}$ in $Z_7$.
	In Step 2, let $D_0,D_1,\ldots, D_6$ be the $7$ lines, where
	$D_0=\{0,1,3\}$, $D_1=\{1,2,4\}$, $D_2=\{2,3,5\}$, $D_3=\{3,4,6\}$,	$D_4=\{4,5,0\}$, $D_5=\{5,6,1\}$, and $D_6=\{6,0,2\}$.
	In Step 3, construct the bipartite graph with the $7$ lines on the left and the $7$ points on the right, and add an edge between a point node and a line node if that point is in that line. This bipartite graph is shown in \rfig{algoex} (a).
	In Step 4, first remove point 0 and line 0 along with the edges attached to these two nodes from the bipartite graph. The nodes and the edges that need to be removed are marked in red in \rfig{algoex} (b), and the trimmed bipartite graph is shown in \rfig{algoex} (c).
	Then, let $G=(g_{n,\ell})$ be the $6 \times 6$ biadjacency matrix of the trimmed bipartite graph with $g_{n,\ell}=1$ if point $n$ is in $D_\ell$, i.e.,
	\beq{ex1111x}
	G ~~=~~ 
	\bordermatrix{     
		& D_1 & D_2 & D_3 & D_4 & D_5 & D_6	\cr
		1 & 1 & 0 & 0 & 0 & 1 & 0 \cr
		2 & 1 & 1 & 0 & 0 & 0 & 1 \cr
		3 & 0 & 1 & 1 & 0 & 0 & 0 \cr
		4 & 1 & 0 & 1 & 1 & 0 & 0 \cr
		5 & 0 & 1 & 0 & 1 & 1 & 0 \cr
		6 & 0 & 0 & 1 & 0 & 1 & 1 \cr
	}
	\eeq
	In Step 5, let $D_{p-a_0}^0 = D_0^0 = \{1,3\}$, $D_{p-a_1}^0 = D_6^0 = \{6,2\}$ and $D_{p-a_2}^0 = D_4^0 = \{4,5\}$ be the $3$ pencil of lines that contain point 0.
	In Step 6, remove the $(p-a_1)$\textsuperscript{th} = 6\textsuperscript{th} and the $(p-a_2)$\textsuperscript{th} = 4\textsuperscript{th} columns in $G$ to form a $6 \times 4$ biadjacency matrix $\tilde G$, i.e.,
	\beq{ex2222}
	G =
	\left ( \begin{array}{llllll}
		1 & 0 & 0 & {\color{blue}0} & 1 & {\color{blue}0} \\
		1 & 1 & 0 & {\color{blue}0} & 0 & {\color{blue}1} \\
		0 & 1 & 1 & {\color{blue}0} & 0 & {\color{blue}0} \\
		1 & 0 & 1 & {\color{blue}1} & 0 & {\color{blue}0} \\
		0 & 1 & 0 & {\color{blue}1} & 1 & {\color{blue}0} \\
		0 & 0 & 1 & {\color{blue}0} & 1 & {\color{blue}1} \\
	\end{array}
	\right )
	\Rightarrow
	\left ( \begin{array}{llll}
		1 & 0 & 0 & 1  \\
		1 & 1 & 0 & 0  \\
		0 & 1 & 1 & 0  \\
		1 & 0 & 1 & 0  \\
		0 & 1 & 0 & 1  \\
		0 & 0 & 1 & 1  \\
	\end{array}
	\right )
	= \tilde{G}
	\eeq
	The two lines that need to be removed are marked in red in \rfig{algoex} (d), and the bipartite graph after removing the two lines are shown in \rfig{algoex} (e).
	In Step 7, let $B=\cup_{i=0}^{d_1-1} D_{\peri-a_i}^0 = D_{p-a_0}^0 = D_0^0 = \{1,3\}$. Then, remove rows of $\tilde G$ that are not in $B$ to form a $2 \times 4$ biadjacency matrix $H$, i.e.,
	\beq{ex3333}
	\tilde{G} = 
	\left ( \begin{array}{llll}
		\color{blue}1 & \color{blue}0 & \color{blue}0 & \color{blue}1  \\
		1 & 1 & 0 & 0  \\
		\color{blue}0 & \color{blue}1 & \color{blue}1 & \color{blue}0  \\
		1 & 0 & 1 & 0  \\
		0 & 1 & 0 & 1  \\
		0 & 0 & 1 & 1  \\
	\end{array}
	\right )
	\Rightarrow
	\left ( \begin{array}{llll}
		\color{blue}1 & \color{blue}0 & \color{blue}0 & \color{blue}1  \\
		\color{blue}0 & \color{blue}1 & \color{blue}1 & \color{blue}0  \\
	\end{array}
	\right )
	=H
	\eeq
	The points in set $B$ along with the edges attached to these nodes are marked in red in \rfig{algoex} (f).
	The output of Algorithm \ref{alg:PPoLs} in this example is the $2 \times 4$ binary pooling matrix $H$.
	\eex
}

\bprop{degree}
The degree of a line node is $d_1$ and the degree of a point node is $m$.
\eprop

\bproof
As the remaining lines are the lines not containing point 0, each line then intersects with $D_{\peri-a_i}^0$ at exactly one point.
Since there are $d_1$ pencil of lines that contain point 0, each line then intersects with $B=\cup_{i=1}^{d_1} D_{\peri-a_i}^0$ at exactly $d_1$ points.
On the other hand, each of the points in $B$ is in a line that contains point 0. As the lines that contain point 0 are removed,
each point in $B$ is in $m$ lines of the remaining $m^2$ lines.
\eproof

\bprop{intersection}
There is at most one {\em common} nonzero element in  two rows (resp. columns) in the pooling matrix $H$ from Algorithm \ref{alg:PPoLs}, i.e., the inner product of two row vectors (resp. column vectors) is at most 1.
\eprop

\bproof
This is because the bipartite graph with the biadjacency matrix $H$ is a subgraph of the line-point bipartite graph corresponding to a finite projective plane.
From (P3) and (P4) of \rdef{fpp}, any two distinct lines intersect at exactly one point, and
any two distinct points lie on exactly one line.
Thus, there is at most one common nonzero element in two rows (resp. columns) in $H$ from Algorithm \ref{alg:PPoLs}.
\eproof

\bcor{girth}
The girth (the minimum length of a cycle) of the bipartite graph with biadjacency matrix $H$ is at least 6.
\ecor

\bproof
As the length of a cycle in a bipartite graph must be an even number. It suffices to show that there does not exist a cycle of length 4.
We prove this by contradiction. Suppose that there is a cycle of length 4. Suppose that this cycle contains two line nodes $L_1$ and $L_2$ and two point nodes $P_1$ and $P_2$. Then the intersection of the two lines $L_1$ and $L_2$ contains two points  $P_1$ and $P_2$. This contradicts (P3) in \rdef{fpp}.
\eproof

\bthe{detection}
Consider using the $d_1 m \times m^2$ pooling matrix $H$ from Algorithm \ref{alg:PPoLs} for a binary state vector $x$
in a noiseless setting. If the number of positive samples in $x$ is not larger than $d_1-1$, then every sample can be correctly decoded by the DD algorithm in Algorithm \ref{alg:binary}.
\ethe

\bproof
Suppose that there are at most $d_1-1$ positive samples. We first show that every negative sample can be correctly decoded by the DD algorithm in Algorithm \ref{alg:binary}.
Consider a negative sample. Since there are at most $d_1-1$ positive samples that can be pooled into the $d_1$ groups of this negative sample, and two different samples can be in a common group at most once (\rprop{intersection}), there must be at least one group without positive samples (among the $d_1$ groups of this negative sample). Thus, this negative sample can be correctly decoded.
Now consider a positive sample. Since there are at most $d_1-2$ positive samples that can be pooled into the $d_1$ groups of this positive sample, and two different samples can be in a common group at most once (\rprop{intersection}), there must be at least one group in which this positive sample is the only positive sample. Thus, every positive sample can be correctly decoded.
\eproof

\begin{figure}[h]
	\centering
	\includegraphics[width=0.4\textwidth]{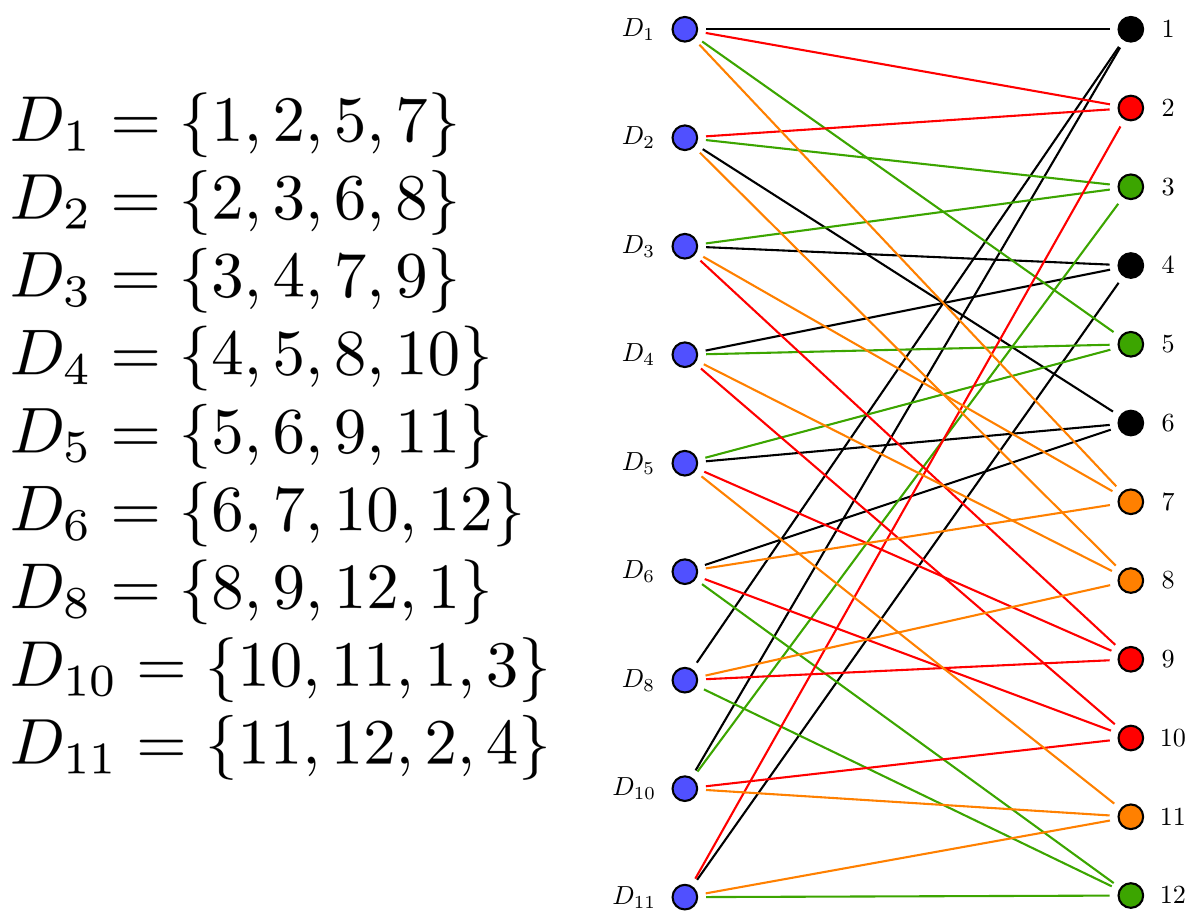}
	\caption{The bipartite graph obtained by using Algorithm \ref{alg:PPoLs} for \rex{map}.}
	\label{fig:eqbi}
\end{figure}

{\color{black}
	\bsubsec{Connection between the PPoL algorithm and the shifted transversal design}{relation}
	
	We note that there are other methods that can also generate bipartite graphs that satisfy the property in \rprop{intersection}. 
	For instance, in the recent paper \cite{taufer2020rapid}, T{\"a}ufer used the shifted transversal design to generate ``mutlipools'' (in Definition 1 of \cite{taufer2020rapid}) that satisfy the property in \rprop{intersection} when $m$ is a prime (in Theorem 3 of \cite{taufer2020rapid}).
	In this section, we establish the connection between the PPoL design and the shift transversal design when $m$ is restricted to a prime. 
	We do this by identifying a mapping between these two designs in the following example.
	
	\bex{map}{}
	Consider $m=3$ in the PPoL algorithm. Then let $p=m^2+m+1=13$, and $D_0 =\{a_0,a_1,a_2,a_3\} = \{0,1,4,6\}$ be a perfect difference set in $Z_{13}$. By using the PPoL algorithm in Algorithm \ref{alg:PPoLs}, we obtain a bipartite graph with 9 samples (lines) and 12 groups (points) in \rfig{eqbi}. In the following, we discuss the four cases with $d_1=1,2,3,4$, respectively.
	
	
	\begin{itemize}
		\item[(i)] If $d_1=1$, $D_{p-a_0}^0 = D_{0}^0=\{1,4,6\}$. Then 
		$D_1, D_{10}, D_{8}$ are in group 1,
		$D_4, D_3, D_{11}$ are in group 4, and
		$D_5, D_2, D_6$ are in group 6.
		Thus, every sample is contained in $d_1=1$ group. (See the black points and lines in \rfig{eqbi}.)
		
		\item[(ii)] If $d_1=2$, $D_{p-a_0}^0 = D_{0}^0=\{1,4,6\}$ and $D_{p-a_1}^0 = D_{12}^0=\{12,3,5\}$. Then, in addition to the pooling results in (i), 
		$D_1, D_4, D_5$ are in group 5,
		$D_{10}, D_3, D_2$ are in group 3, and
		$D_8, D_{11}, D_6$ are in group 12.
		Thus, every sample is contained in $d_1=2$ groups. (See the black and green ones in \rfig{eqbi}.)
		
		\item[(iii)] If $d_1=3$, $D_{p-a_0}^0 = D_{0}^0=\{1,4,6\}$, $D_{p-a_1}^0 = D_{12}^0=\{12,3,5\}$, and $D_{p-a_2}^0 = D_9^0=\{9,10,2\}$. Then, in addition to the pooling results in (i) and (ii),
		$D_8, D_3, D_5$ are in group 9,
		$D_{10}, D_4, D_6$ are in group 10, and
		$D_1, D_{11}, D_2$ are in group 2.
		Thus, every sample is contained in $d_1=3$ groups. (See the black, green, and red ones in \rfig{eqbi}.)

		\item[(iv)] If $d_1=4$, $D_{p-a_0}^0 = D_{0}^0=\{1,4,6\}$, $D_{p-a_1}^0 = D_{12}^0=\{12,3,5\}$, $D_{p-a_2}^0 = D_9^0=\{9,10,2\}$, and $D_{p-a_3}^0 = D_7^0 = \{7,8,11\}$. Then, in addition to the pooling results in (i), (ii) and (iii),
		$D_1, D_3, D_6$ are in group 7,
		$D_8, D_4, D_2$ are in group 8, and
		$D_5, D_{10}, D_{11}$ are in group 11.
		Thus, every sample is contained in $d_1=4$ groups. (See the black, green, red, and orange ones in \rfig{eqbi}.)
	\end{itemize} 
	The above PPoL pooling strategy is the same as $(N,n,k)=(m^2,m,d_1)$-multipool in the shifted transversal design \cite{taufer2020rapid} if we arrange the 9 samples in the $3\times 3$-square in Table \ref{tab:multipool}. 
	Specifically, pooling along rows yields the three groups $\{D_1, D_{10}, D_8\}$, $\{D_4, D_3, D_{11}\}$, and $\{D_5, D_2, D_6\}$. This corresponds to the case with $d_1=1$ in the PPoL design. On the other hand, pooling along columns yields the three groups $\{D_1, D_4, D_5\}$, $\{D_{10}, D_3, D_2\}$, and $\{D_8, D_{11}, D_6\}$. This corresponds to the case with $d_1=2$ in the PPoL design. Moreover, pooling with slope 1 (resp. 2) corresponds to the case with $d_1=3$ (resp. $d_1=4$).
	
	\begin{table}[h]
		\centering
		\caption{Arrangement of the 9 samples in a $3\times 3$ rectangular grid.}
		\label{tab:multipool}
		\begin{tabular}{|l|l|l|}
			\hline
			$D_1$ & $D_{10}$ & $D_8$ \\ \hline
			$D_4$ & $D_3$ & $D_{11}$ \\ \hline
			$D_5$ & $D_2$ & $D_6$ \\ \hline
		\end{tabular}
	\end{table}
	
	In fact, these two constructions are closely related to orthogonal Latin squares \cite{Euler}. 
	For $n=3$ (which is a prime power), there are exactly $n-1=2$ mutually orthogonal Latin squares: $\{C^{(r)}=c_{i,j}^{(r)}: r=1,2\}$, where $c_{i,j}^{(r)}=(r * i+j)$ is in GF(3).
	With the ``vertical'' and ``horizontal'' cases, the maximum number of multiplicity $k$ in the shifted transversal design is $n+1=4$. Similarly, the maximum number of $d_1$ in the PPoL algorithm is $m+1=4$.
	Moreover, pooling matrices that satisfy the decoding property in \rthe{detection} are known as the superimposed codes in \cite{kautz1964nonrandom}.
	\eex
}

\ \\

\bsubsec{Probabilistic analysis of the PPoL pooling matrices}{analysis}


In this section, we conduct a probabilistic analysis of the PPoL pooling matrices.
We make the following assumption:
\begin{description}
	\item[(A1)] All the samples are i.i.d. Bernoulli random variables. A sample is positive (resp. negative) with probability
	$r_1$ (resp. $r_0$). The probability $r_1$ is known as the prevalence rate in the literature.
\end{description}
Note that $r_1+r_0=1$. Also, let $q_1$ (resp. $q_0$) be the probability that the  group end of a randomly selected edge is positive (resp. negative). Excluding the randomly selected edge,  there are $d_2-1$ remaining edges in that group, and thus
\bear{groupt3333}
q_0&=&(r_0)^{d_2-1}, \\
q_1&=&1-(r_0)^{d_2-1}.
\eear

\begin{figure}
	\centering
	\includegraphics[width=0.35\textwidth]{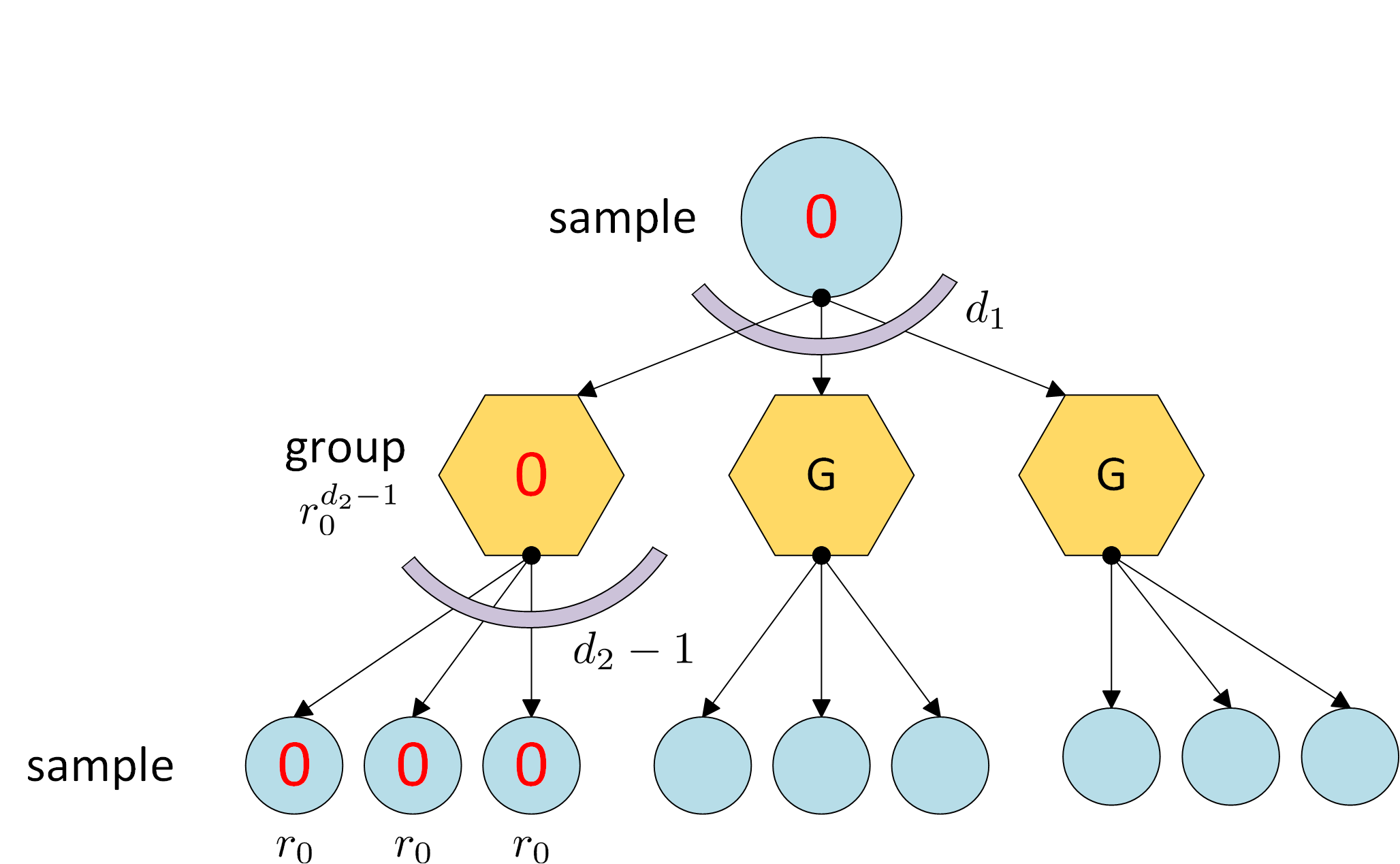}
	\caption{Computing the conditional probability $p_0$ by the tree evaluation method.}
	\label{fig:prob0}
\end{figure}

Let $p_0$ be the conditional probability that a  sample {\em cannot} be decoded, given that the sample is a negative sample. Note that a negative sample can be decoded if at least one of its edges is in a negative group, excluding its edge (see \rfig{prob0}).
Consider a negative sample, called the tagged sample. Since the girth of the bipartite graph of the pooling matrix is 6 (as shown in \rcor{girth}), the
samples in the $d_1$ groups of the subtree of the tagged sample are distinct (see the tree expansion in \rfig{prob0}). Thus,
\beq{groupt4444}
p_0=(q_1)^{d_1}=(1-(r_0)^{d_2-1})^{d_1}.
\eeq

\begin{figure}
	\centering
	\includegraphics[width=0.35\textwidth]{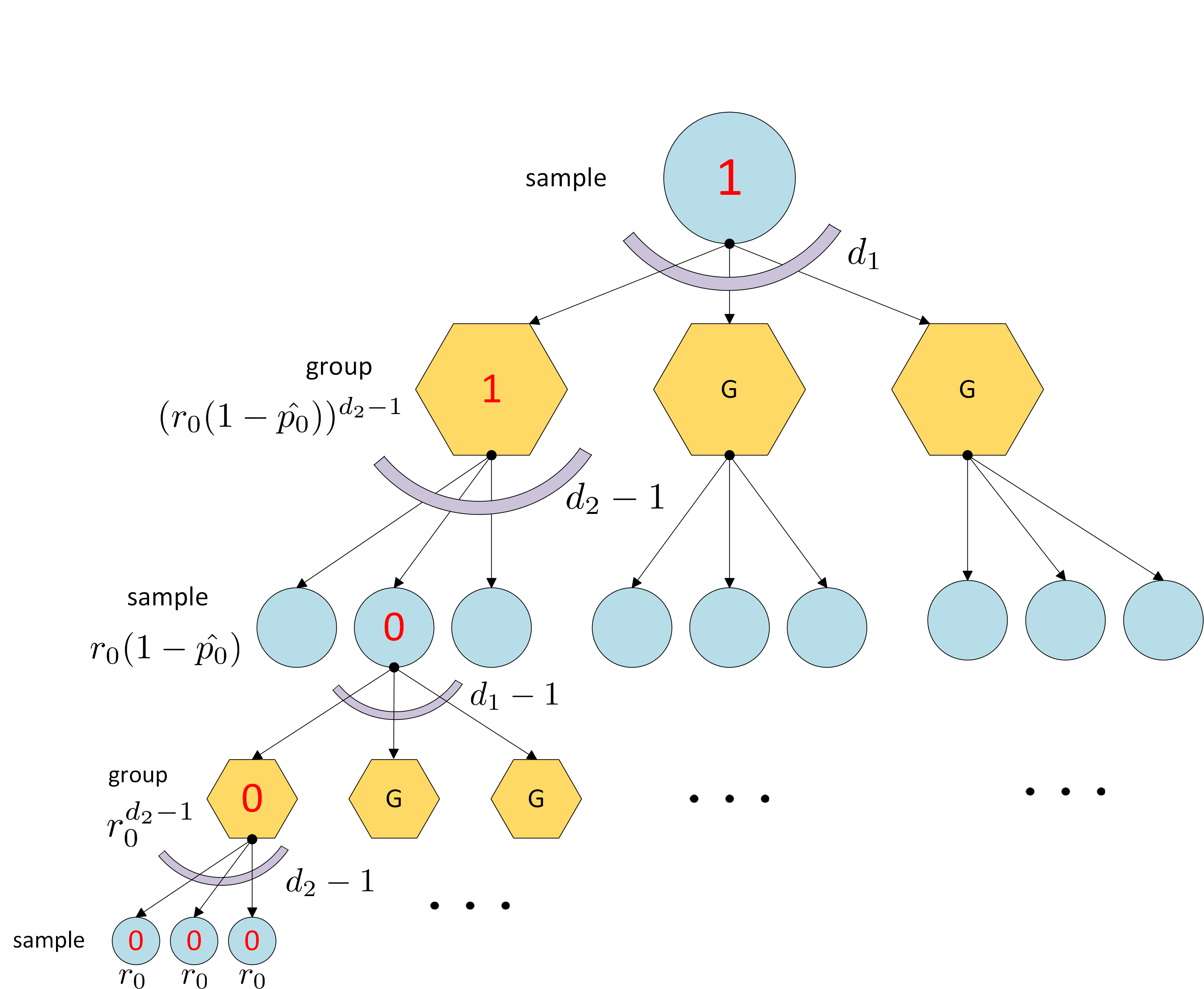}
	\caption{Computing the conditional probability $p_1$ by the tree evaluation method.}
	\label{fig:prob1}
\end{figure}

Let $\hat p_0$ be the conditional probability that the sample end of a randomly selected edge {\em cannot} be decoded, given that the sample end is a negative sample. Note that the excess degree of a sample (excluding the randomly selected edge)
is $d_1-1$. Analogous to the argument for \req{groupt4444} (see the bottom subtree of the tree expansion in \rfig{prob1}), we have
\beq{groupt4444excess}
\hat p_0=(q_1)^{d_1-1}=(1-(r_0)^{d_2-1})^{d_1-1}.
\eeq

Let $p_1$ be the conditional probability that a  sample {\em cannot} be decoded given that the sample is a positive sample. Note that a positive sample can be decoded if at least one of its edges is in a group in which all the edges are removed except the edge of the positive sample.
Since an edge is removed if its sample end is a negative sample and that sample end is decoded to be negative,
the probability that an edge is removed is $(1-\hat p_0) r_0$.
If the tree expansion in \rfig{prob1} is actually a tree,  then
\beq{groupt5555}
p_1=(1-(r_0 (1-\hat p_0))^{d_2-1})^{d_1}.
\eeq
We note that the tree expansion in \rfig{prob1} may {\em not} be  a tree for a PPoL pooling matrix generated from Algorithm \ref{alg:PPoLs}, the identity in \req{groupt5555} is only an approximation. 
A sufficient condition for the tree expansion in \rfig{prob1} to be a tree of depth 4 is that the girth of the bipartite graph is larger than 8.
(If the graph in \rfig{prob1} is not a tree, i.e., there is a loop in that graph, then the girth of the bipartite graph is less than or equal to 8.)
Unfortunately, the girth of a PPoL pooling matrix can only proved to be at least 6.
Since a sample cannot be decoded with probability $r_0 p_0+r_1 p_1$, the average number of tests needed for the DD2 algorithm in Algorithm \ref{alg:binary2} to decode the $M$ samples is $N+M(r_0 p_0+r_1 p_1)$.
The expected relative cost for the DD2 algorithm with an $N \times M$ pooling matrix is
\beq{groupt7777}
\frac{N+M(r_0 p_0+r_1 p_1)}{M}=\frac{1}{G}+{r_0 p_0+r_1 p_1},
\eeq
where $G=M/N$ is the (compressing) gain of the pooling matrix in \req{groupt2222}.
Note that for a $(d_1, d_2)$-regular pooling matrix, we have from \req{groupt2222} that $G={d_2}/{d_1}$.
Thus, we can use \req{groupt4444}, \req{groupt5555} and \req{groupt7777} to find the
$(d_1, d_2)$-regular pooling matrix that has the lowest expected relative cost (though \req{groupt5555} is only an approximation for the pooling matrices constructed from the PPoL algorithm). In Table \ref{tab:minimum}, we
use grid search to find the $(d_1, d_2)$-regular pooling matrix with the lowest expected relative cost for various prevalence rates $r_1$ up to 10\%. The search regions for the grid search are $2 \le d_1 \le 8$ and $d_1 \le d_2 \le 31$.
In the last column of this table, we also show the expected relative cost of the Dorfman two-stage algorithm (Table I of \cite{dorfman1943detection}).
As shown in this table, using the DD2 algorithm (with the optimal pooling matrices) has significant gains over the Dorfman two-stage algorithm.
Unfortunately, not every optimal $(d_1,d_2)$-regular pooling matrix in Table \ref{tab:minimum} can be constructed by using the PPoL algorithm in Algorithm \ref{alg:PPoLs}. In the next section, we will look for suboptimal pooling matrices that have small performance degradation.

\begin{table}[ht]
	\centering
	\caption{The $(d_1, d_2)$-regular pooling matrix with the lowest expected relative cost from \req{groupt7777}.}
	\label{tab:minimum}
	\begin{tabular}{|c|c|c|c|c|}
		\hline
		$r_1$ & $d_1$ & $d_2$ & cost \req{groupt7777} & Dorfman \cite{dorfman1943detection}\\ \hline
		1\% & 3 &  31 &  0.1218 & 0.20 \\ \hline
		2\% & 4 &  29 &  0.1881  & 0.27 \\ \hline
		3\% & 4 &  22 &  0.2545 & 0.33 \\ \hline
		4\% & 4 &  17 &  0.3147  & 0.38 \\ \hline
		5\% & 3 &  12 &  0.3678 & 0.43 \\ \hline
		6\% & 3 &  11 &  0.4166 & 0.47 \\ \hline
		7\% & 3 &  10 &  0.4627 & 0.50 \\ \hline
		8\% & 2 &  7 &  0.5035 & 0.53 \\ \hline
		9\% & 2 &  6 &  0.5416 &0.56\\ \hline
		10\% & 2 &  6 &  0.5760  &0.59\\ \hline
	\end{tabular}
\end{table}

\bsec{Numerical Results}{num}

In this section, we compare the performance of various pooling matrices by using the DD2 algorithm in Algorithm \ref{alg:binary2}.
The first four pooling matrices are constructed by using the PPoL algorithm in Algorithm \ref{alg:PPoLs} with the parameters $(m,d_1)=(31,3)$, $(23,4)$, $(13,3)$, and $(7,2)$, respectively.  The fifth pooling matrix is the pooling matrix used in P-BEST \cite{shental2020efficient}. The sixth matrix is the $15 \times 35$ pooling matrix constructed by the Kirkman triples.
The next two pooling matrices are used in Tapestry \cite{ghosh2020tapestry,ghosh2020compressed}. The last pooling matrix is the 2D-pooling matrix in \cite{sinnott2020evaluation}.
In Table \ref{tab:basic_info}, we show the basic information of these pooling matrices.
The size of an $N\times M$ pooling matrix indicates that the number of groups is $N$, and the number of samples is $M$.
The parameter $d_1$ is the number of groups in which a sample is pooled.
On the other hand, $d_2$ is the number of samples in a group. Note that there are some pooling matrices that are not $(d_1,d_2)$-regular. For instance, in the 2D-pooling matrix, there are 8 groups with 12 samples, and 12 groups with 8 samples. Also, both the $16 \times 40$ matrix and the $24 \times 60$ matrix used in Tapestry are not $(d_1,d_2)$-regular.
The column marked with {\em row cor.} (resp. {\em col. cor.}) is
the maximum of the inner product of two rows (resp. columns) in a pooling matrix.
For a pooling matrix,  the column marked with {\em girth} is the minimum length of a cycle in the bipartite graph corresponding to that pooling matrix.
The column marked with {\em (comp.) gain} is the compressing gain $G$ of a pooling matrix, which is  the ratio of the number of columns (samples) to the number of rows (groups), i.e., $G=M/N$.
As shown in Table \ref{tab:basic_info}, both the row correlation and the column correlation of the pooling matrices constructed from
the PPoL algorithm in Algorithm \ref{alg:PPoLs} are 1. So are the $15 \times 35$ pooling matrix constructed by the Kirkman triples. Such a correlation result is expected from \rprop{intersection}. On the other hand, the row correlation
and the column correlation of the pooling matrix in P-BEST \cite{shental2020efficient} are 6 and 2, respectively.  Also, the girth of
the pooling matrix in P-BEST  is only 4, which is smaller than the other four matrices.
The girth of the $16\times 40$ pooling matrix in Tapestry is also 4.
This shows that the pooling matrices from the PPoL algorithm are more ``spread-out'' than the pooling matrix in P-BEST and the $16\times 40$ pooling matrix in Tapestry.

\begin{table}[ht]
	\centering
	\caption{Basic information of some pooling matrices.}
	\label{tab:basic_info}
	\begin{tiny}
		\begin{tabular}{|c|c|c|c|c|c|c|c|}
			\hline
			$H$ & size & $d_1$ & $d_2$ & \begin{tabular}[c]{@{}c@{}}row\\ cor.\end{tabular} & \begin{tabular}[c]{@{}c@{}}col.\\ cor.\end{tabular} & girth & \begin{tabular}[c]{@{}c@{}}(comp.)\\ gain\end{tabular}   \\ \hline
			PPoL-(31,3) & $ 93 \times 961 $ & 3 & 31 & 1 & 1 & 6 & 10.33 \\ \hline
			PPoL-(23,4) & $ 92 \times 529 $ & 4 & 23 & 1 & 1 & 6 & 5.75  \\ \hline
			PPoL-(13,3) & $ 39 \times 169 $ & 3 & 13 & 1 & 1 & 6 & 4.33  \\ \hline
			PPoL-(7,2)  & $ 14 \times 49  $ & 2 & 7  & 1 & 1 & 8 & 3.5   \\ \hline
			P-BEST Matrix \cite{shental2020efficient} & $48 \times 384$ & 6 & 48 & 6 & 2 & 4 & 8 \\ \hline
			Kirkman Matrix $15\times 35$  & $15 \times 35$ & 3 & 7 &1 & 1 & 6 & 2.33 \\ \hline
			Tapestry Matrix $16\times 40$ \cite{ghosh2020tapestry} & $16 \times 40$ & 2-3 & 6-9 &3 & 2 & 4 & 2.5 \\ \hline
			Tapestry Matrix $24 \times 60$ \cite{ghosh2020tapestry} & $24 \times 60$ & 2-3 & 6-7 &1 & 1 & 6 & 2.5 \\ \hline
			2D-pooling Matrix \cite{sinnott2020evaluation} & $20 \times 96$ &2 & 12(8) & 1 & 1 & 8 & 4.8 \\ \hline
		\end{tabular}
	\end{tiny}
\end{table}

\begin{figure}
	\centering
	\includegraphics[width=0.45\textwidth]{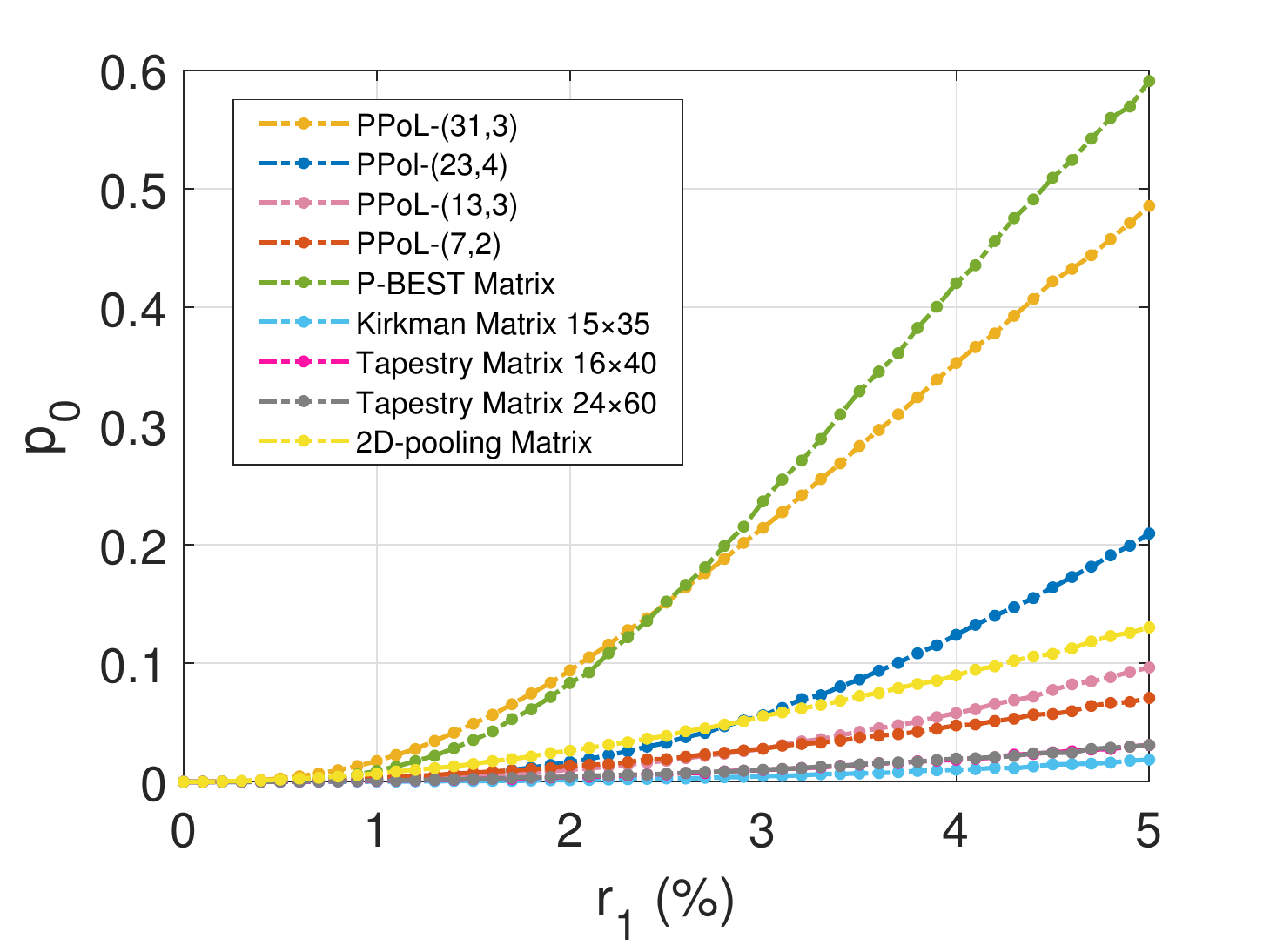}
	\caption{The conditional probability $p_0$  (that a sample cannot be decoded given it is a {\em negative} sample) as a function of the prevalence rate $r_1$ for various pooling matrices.}
	\label{fig:comparep0}
\end{figure}

To compare the performance of these pooling matrices, we conduct 10,000 independent experiments for each value of the prevalence rate $r_1$, ranging from $0\%$ to $5\%$. Each numerical result is obtained by averaging over these 10,000
independent experiments.
In \rfig{comparep0}, we show the (measured) conditional probability $p_0$  (that a sample cannot be decoded given it is a {\em negative} sample) for these pooling matrices. For the PPoL pooling matrices, the measured $p_0$'s match extremely well with the theoretical results from \req{groupt4444}.
As shown in this figure, the Kirkman matrix and the two matrices in Tapestry have the best performance. This is because  their $d_2$'s (the number of samples in a group) are small (below 9 for these three matrices).  As such, the probability that a group is tested negative is higher than the other pooling matrices. Note that these three matrices also have low (compressing) gains, 2.33-2.5.
On the other hand, P-BEST has the worst performance for $p_0$ as the number of samples in a group for that matrix is 48, which is the largest among all these pooling matrices.

\begin{figure}
	\centering
	\includegraphics[width=0.45\textwidth]{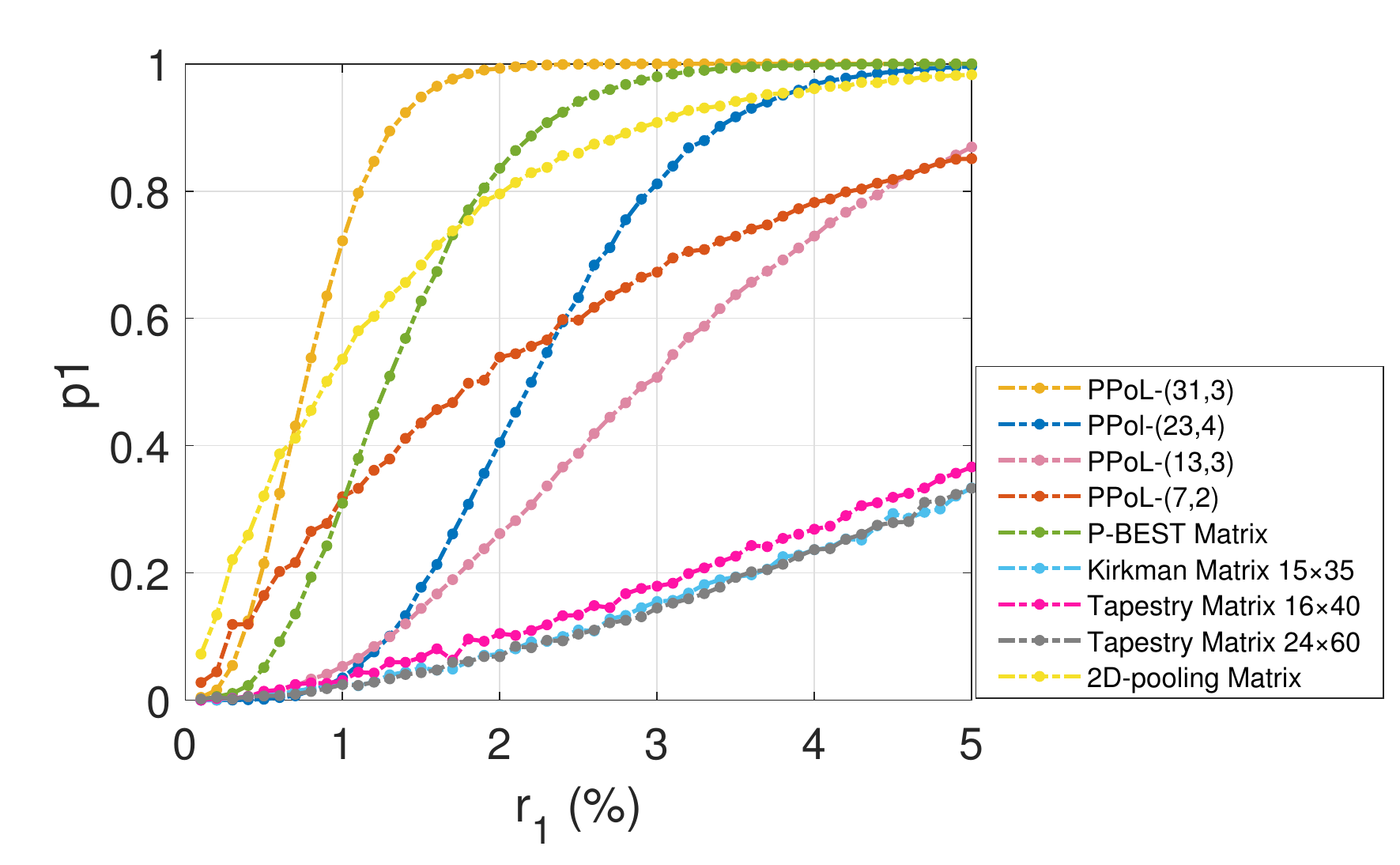}
	\caption{The conditional probability $p_1$  (that a sample cannot be decoded given it is a {\em positive} sample) as a function of the prevalence rate $r_1$ for various pooling matrices.}
	\label{fig:comparep1}
\end{figure}

In \rfig{comparep1}, we show the (measured) conditional probability $p_1$  (that a sample cannot be decoded given it is a {\em positive} sample) for these pooling matrices. Once again, the Kirkman matrix and the two matrices in Tapestry have the best performance. This is mainly due to the low (compressing) gains of these three matrices.
Though not shown in \rfig{comparep1}, we note that  the measured $p_1$'s are very close to those from \req{groupt5555}, and thus the tree expansion in  \rfig{prob1} is actually tree-like.

\begin{figure}
	\centering
	\includegraphics[width=0.45\textwidth]{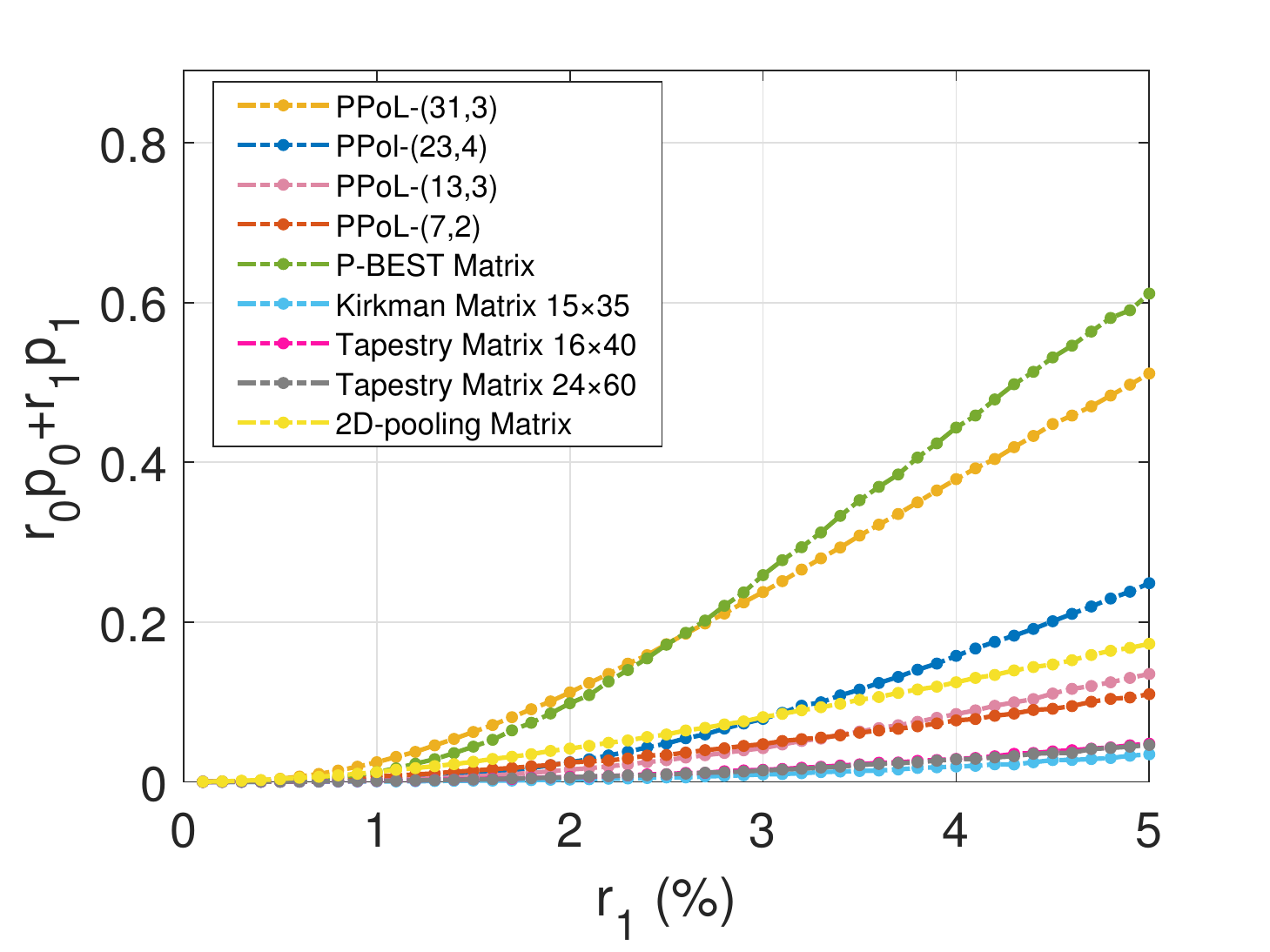}
	\caption{The probability $r_0p_0+r_1p_1$ (that a sample cannot be decoded at the first stage and should be tested individually at the second stage) as a function of the prevalence rate $r_1$ for various pooling matrices.}
	\label{fig:numsec}
\end{figure}

As discussed in \rsubsec{analysis}, the probability that a sample cannot be decoded is $r_0 p_0+r_1 p_1$. Such a probability is also the probability that a sample needs to go through the second stage for individual testing.
In \rfig{numsec}, we show the probability $r_0p_0+r_1p_1$  as a function of the prevalence rate $r_1$ for various pooling matrices.
As shown in this figure, the Kirkman matrix and the two matrices in Tapestry have the best performance. Once again,
this is mainly due to the low (compressing) gains of these three matrices. We note that it takes time to do the second test. The numerical results in \rfig{numsec} imply that using the Kirkman matrix (or the two matrices in Tapestry) has the shortest expected time to obtain a testing result.

\begin{figure}
	\centering
	\includegraphics[width=0.45\textwidth]{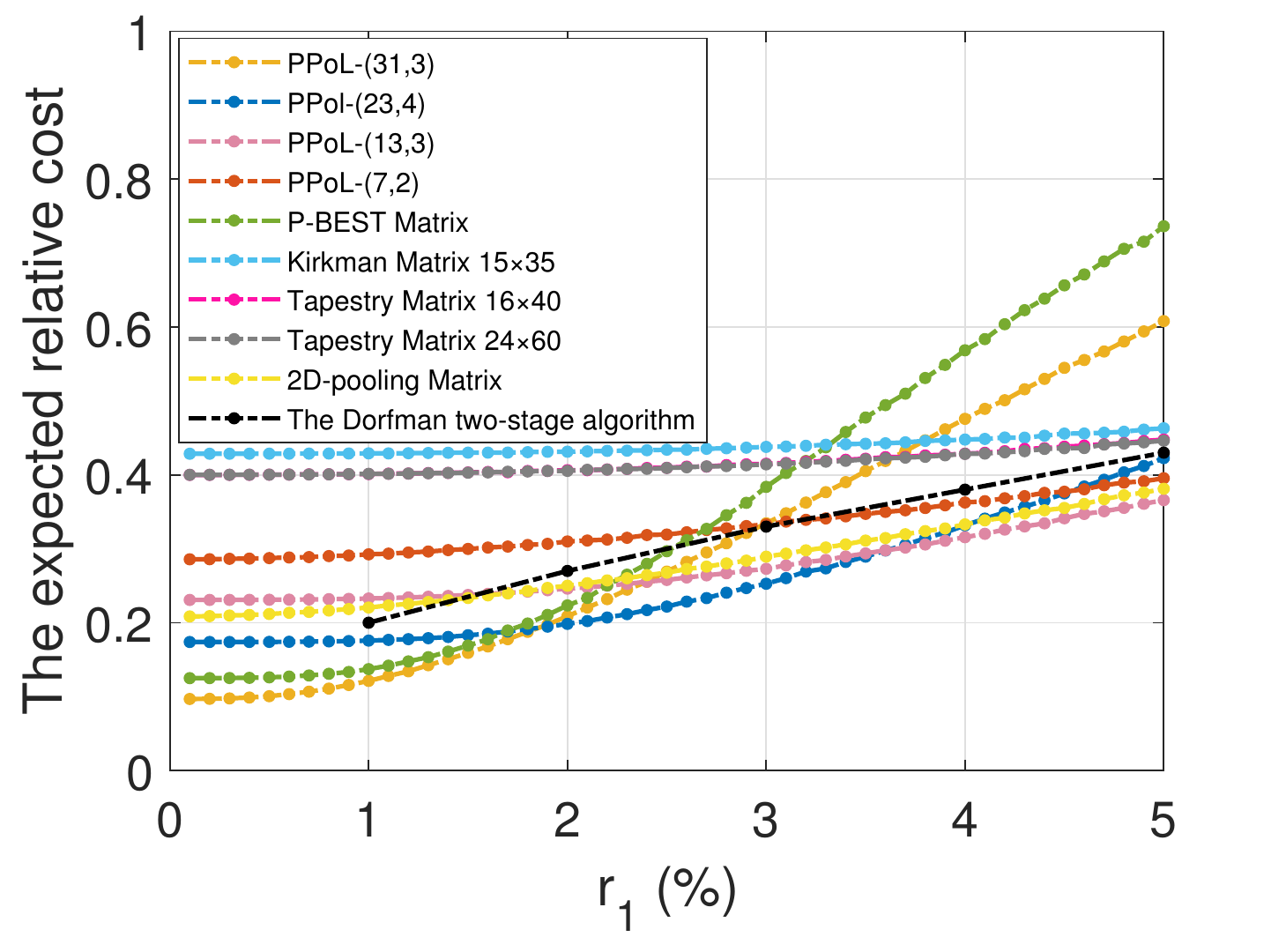}
	\caption{The expected relative cost as a function of the prevalence rate $r_1$ for various pooling matrices.}
	\label{fig:relative}
\end{figure}

A fair comparison of these pooling matrices is to measure their expected relative costs (defined in \cite{dorfman1943detection}). Recall that
the expected relative cost is the ratio of the expected number of tests required by the group testing technique to the number of tests required by the individual testing. In \rfig{relative}, we show the (measured) expected relative costs for these pooling matrices. In this figure, we also plot the curve for the Dorfman two-stage algorithm (the black curve) with the optimal group size $M$ chosen from Table 1 of \cite{dorfman1943detection} for the prevalence rates, $1\%, 2\%, \ldots, 5\%$. To our surprise, the curves for the Kirkman matrix and the two matrices in Tapestry are above the black curve. This means that the expected relative costs of these three matrices are higher than the (optimized) Dorfman two-stage algorithm. Thus, if the additional amount of time to go through the second stage is not critical, using other pooling matrices could lead to more cost reduction than using these three matrices. There are several pooling matrices that have very low relative costs when the prevalence rates are below 1\%. The P-BEST pooling matrix is one of them. However, the relative cost of the P-BEST pooling matrix increases dramatically when the prevalence rates are above 1.3\%. Moreover, the P-BEST pooling matrix has a higher relative cost than the (optimized) Dorfman two-stage algorithm when the prevalence rate is above 2.5\%. On the other hand, 2D-pooling has a very low relative cost when the prevalence rates are above 2.5\%.
To summarize, there does not exist a pooling matrix that has  the lowest relative cost in the whole range of the prevalence rates considered in our experiments.

To optimize the performance, one should choose the right pooling matrix, depending on the prevalence rate.
However, this might be difficult as the exact prevalence rate of a new outbreak of COVID-19 in a region might not be known in advance. Our suggestion is to use suboptimal PPoL matrices for a range of prevalence rates, as shown in Table \ref{tab:suboptimal}.
As shown in this table, the costs computed from the theoretical approximations in \req{groupt7777}
and the costs measured from simulations are very close, and they are within 2\% of the minimum costs for $(d_1,d_2)$-regular pooling matrices in Table \ref{tab:minimum}.
From our numerical results in \rfig{relative}, we suggest using the PPoL matrix with $d_1=3$ and $d_2=31$ when the prevalence rate $r_1$ is below 2\%.
In this range of prevalence rates, its expected relative cost is even smaller than that of P-BEST. Moreover, it can achieve an 8-fold reduction in test costs when the prevalence rate is near 1\% (as shown in Table \ref{tab:suboptimal}), and most samples can be decoded in the first stage (as shown in \rfig{numsec}).
When the prevalence rate $r_1$ is between 2\%-4\%, we suggest using the PPoL matrix with $d_1=4$ and $d_2=23$.  In this range of prevalence rates, using such a pooling matrix can still
achieve (at least) a 3-fold reduction in test costs. Roughly, 17\% of samples need to go through the second stage when the prevalence rate is near 4\% (as shown in \rfig{numsec}).
When the prevalence rate $r_1$ is between 4\%-7\%,
we suggest using the PPoL matrix with $d_1=3$ and $d_2=13$, and it can still achieve
(at least) a 2-fold reduction in test costs.
When the prevalence rate $r_1$ is between 7\%-10\%,
we suggest using the PPoL matrix with
$d_1=2$ and $d_2=7$. Though its expected relative cost is still lower than that of the Dorfman two-stage algorithm, the difference is small.

\begin{table}[ht]
	\centering
	\caption {Suboptimal PPoL pooling matrices.}
	\label{tab:suboptimal}
	\begin{tabular}{|c|c|c|c|c|c|}
		\hline
		$r_1$ &  $d_1$ & $d_2$	& cost \req{groupt7777} &	cost (sim) & Dorfman \cite{dorfman1943detection} \\ \hline
		1\%	& 3	& 31 &	0.1218 & 0.12 &	0.20 \\ \hline
		2\%	& 4	& 23 &	0.1973 & 0.20 &	0.27 \\ \hline
		3\%	& 4	& 23 &	0.2552 & 0.25 &	0.33 \\ \hline
		4\%	& 3	& 13 &	0.3170 & 0.32 & 0.38 \\ \hline
		5\%	& 3	& 13 &	0.3685 & 0.37 &	0.43 \\ \hline
		6\%	& 3	& 13 &	0.4243 & 0.42 &	0.47 \\ \hline
		7\%	& 2	& 7	 &  0.4651 & 0.47 &	0.50 \\ \hline
		8\%	& 2	& 7	 &  0.5035 & 0.50 &	0.53 \\ \hline
		9\%	& 2	& 7	 &  0.5422 & 0.54 &	0.56 \\ \hline
		10\% &2	& 7	 &  0.5809 & 0.58 & 0.59 \\ \hline
	\end{tabular}
\end{table}

\bsec{Conclusion}{con}

In this paper, we proposed a new family of PPoL polling matrices that have maximum column correlation and row correlation of 1 for a wide range of
column weights.
Using the two-stage definite defectives (DD2) decoding algorithm,
we compare their performance with various pooling matrices
proposed in the literature, including 2D-pooling \cite{sinnott2020evaluation}, P-BEST \cite{shental2020efficient}, and Tapestry \cite{ghosh2020tapestry,ghosh2020compressed}. Our numerical results showed no pooling matrix with the lowest expected relative cost in the whole range of the prevalence rates. To optimize the performance, one should choose the right pooling matrix, depending on the prevalence rate. As the family of PPoL matrices can dynamically adjust their construction parameters according to the prevalence rates, it seems that using such a family of pooling matrices might lead to better cost reduction than using a fixed pooling matrix.

There are several research directions for future works:
\begin{description}
	\item[(i)] Other decoding algorithms: in this paper, we only evaluated the performance of pooling matrices using the DD2 algorithm. To probe further, we are currently investigating the possibility of using other decoding algorithms, in particular, the GPSR algorithm in \cite{shental2020efficient} and the belief propagation (BP) algorithm in \cite{sejdinovic2010note}.
	\item[(ii)] Noisy decoding: The DD2 algorithm works very well in the noiseless setting. However, it is not clear whether it can continue to perform well in a noisy setting. There are several noise models proposed in the literature (see, e.g., the monograph \cite{aldridge2019group}). Among them, the dilution noise model is of particular interest to us. Our preliminary numerical results show that the number of false negatives of the DD2 algorithm might increase significantly with respect to the increase of the dilution noise. As such, one should be cautious about using the DD2 algorithm when the dilution noise is not negligible.
	{\color{black}
		Another recent work \cite{zhu2020noisy} deals with noisy pooled testing, where a noisy communication channel causes false positives and false negatives.  
		To decode samples, the authors of \cite{zhu2020noisy} proposed the generalized approximate message passing (GAMP) algorithm that requires both false-negative and false-positive probabilities.
		But the exact channel conditions, including false-negative and false-positive probabilities, are very difficult to estimate in practice. 
	}
	\item[(iii)] Ternary samples: in this paper, we only considered binary samples. For ternary samples, there are three test outcomes: negative (0), weakly positive (1), and strongly positive (2). It seems possible to extend the DD2 algorithm for binary samples to the setting with ternary samples by using successive cancellations.
\end{description}

\end{document}